\documentclass[aps,pre,reprint,twocolumn,showpacs]{revtex4-1}
\usepackage{graphicx,amssymb,amsmath}

\begin{document}

\title{Branching and annihilating random walks: exact results at low branching rate}

\author{Federico Benitez} 
\affiliation{LPTMC, CNRS-UMR 7600, Universit\'e Pierre et Marie Curie, 75252 Paris, France}
\author{Nicol\'as Wschebor} 
\affiliation{Instituto de F\'{\i}sica, Facultad de Ingenier\'{\i}a, Universidad de la Rep\'ublica, 
J.H.y Reissig 565, 11000 Montevideo, Uruguay}
\date{\today}

\begin{abstract}
We present some exact results on the behavior of Branching and Annihilating Random Walks, both in the Directed Percolation and Parity Conserving universality classes. Contrary to usual perturbation theory, we perform an expansion in the branching rate around the non trivial Pure Annihilation model, whose correlation and response function we compute exactly. With this, the non-universal threshold value for having a phase transition in the simplest system belonging to the Directed Percolation universality class is found to coincide with previous Non Perturbative Renormalization Group approximate results. We also show that the Parity Conserving universality class has an unexpected RG fixed point structure, with a PA fixed point which is unstable in all dimensions of physical interest.
\end{abstract}

\maketitle

\section{Introduction}

The study of critical behavior in out of equilibrium systems has been a very active topic in statistical mechanics during the last decades \cite{janssen79,hinrichsen00,henkel}. As in equilibrium, fluctuations and correlations become large in systems close to a continuous phase transition, leading to divergences in quantities such as correlation time and length, and to emergent phenomena classifiable (as in equilibrium) in different universality classes. 

Renormalization Group (RG) methods have been employed since their development to the study of critical dynamics \cite{janssen79}. These are well suited for the task, given that the diverging correlation length in second order phase transitions signals the emergence of scale free behavior, whereas the RG approach focusses on how systems change under scale transformations. 

In this work we study systems which attain a non-thermal (non-equilibrium) steady state at long times, having a stationary probability distribution which is not constrained by a detailed balance condition. These out of equilibrium system usually exhibit a much richer variety of phenomena than their counterparts at or close to equilibrium. In particular, usual RG techniques have to be adapted, given that one does not know the explicit probability distribution, in the way one knows the Boltzmann-Gibbs distribution when in equilibrium.

Perturbative RG has been used in the study of second order out of equilibrium phase transitions, although not with the same level of success as in the study of equilibrium phase transitions. This is due in part to the absence of high-order perturbative results, as opposed to equilibrium problems, but also to deeper physical issues. For example, for most of these systems, upper critical dimensions happen to be generally far from the dimensions of physical interest, complicating the usual $\epsilon$-expansion. Moreover, these models generally lack a lower critical dimension or (generally speaking) an exactly solvable low $d$ model. Finally, out of equilibrium systems also tend to be more prone to show genuinely non-perturbative behavior, such as large couplings \cite{Delamotte:ito,BMWkpz,BMWkpz2}.

Here we study some properties of phase transitions occurring in what are known as Branching and Annihilating Random Walks (BARW) \cite{henkel,grassberger84,cardy96,tauber05,janstauber}, that is, systems composed of particles of a single species $A$, which diffuse in a $d$-dimensional space, and which can suffer both annihilation and branching (i.e. offspring creation) processes, with different rates. From these competing processes typically emerges, at long times, a stationary state which can either be in an active or an absorbing phase, with the absorbing phase corresponding to a no-particles, no-fluctuations state. The existence of this absorbing state implies in particular the absence of detailed balance (and even ergodicity) in these systems. The transition between both phases, which can take place depending on the microscopic rates, is typically of a continuous type. BARW are not only of direct physical interest, but also present a relatively simple class of out of equilibrium systems, and have shown to be very useful for the study of the role of fluctuations in out of equilibrium statistical physics \cite{henkel}.

Due to universality, it is in general enough to consider the simplest possible reactions, such as for example 
\begin{equation}\label{defodd}
2A \xrightarrow{\lambda} \emptyset, \qquad A \xrightarrow{\sigma} 2A 
\end{equation}
or
\begin{equation}\label{defeven}
2A \xrightarrow{\lambda} \emptyset, \qquad A \xrightarrow{\sigma} 3A
\end{equation}
(where $A\to 2A$ does not exist in the second case) as these reactions are the most relevant in the RG sense. At the mean field level (that is, for the classical rate equations) no phase transition is found. That is, fluctuations are here responsible for the very existence of a phase transition. This is in stark contrast with most other known phase transitions, where mean field results predict in general the presence or absence of a phase transition, even when they are unable to yield accurate phase diagrams or critical exponents. In the case of BARW, the mean field result shows the need for taking into account statistical fluctuations, a task for which one expects RG methods to be particularly effective. Notice that in the definition of BARW we exclude explicitly the reaction $A\to \emptyset$. If such reaction is present, the phase diagram is qualitatively well described at a mean field level. Notice also that in this work we will not be dealing with the Pair Contact Processes with Diffusion (PCPD) 
universality class \cite{pcpd,chate}, which can be seen, in terms of BARW, as systems whose reactions involve always necessarily at least two particles.
 
BARW can be classified into sub-classes \cite{henkel,cardy96}, and in this work we will concentrate on the simplest two, which depend on the presence or absence of a symmetry conserving the parity of the number of particles. If no such symmetry exists, as is the case of the system defined by (\ref{defodd}), it has been shown that the BARW system belongs to the Directed Percolation (DP) universality class \cite{janssen81}, whenever a second order phase transition takes place. When only reactions preserving the parity of the number of particles (e.g. the system (\ref{defeven})) are present, an additional symmetry appears, changing the universal properties of the system. If a phase transition takes place in this case, it is known to be in the Parity Conserving (PC) universality class \cite{henkel} (also more properly known as Generalized Voter universality class \cite{chate}). From now on we will refer to these systems as BARW-DP and BARW-PC respectively. 

There exist various known results about BARW, and even some exact results for low-order vertices \cite{peliti} or for special BARW systems which do not present phase transitions \cite{mobilia}. Within perturbative RG, a phase transition for the simplest BARW-DP system (the one consisting of the reactions (\ref{defodd})) is found for space dimensions $d \leq 2$ only \cite{cardy96}. This improves the mean field result, but still contradicts Monte-Carlo and Non Perturbative Renormalization Group (NPRG) results, which observe a phase transition for any $d$ \cite{canet2,canet1}. This difficulty of the perturbative approach may have to do with the fact that for $d>2$ the transition occurs for values of the annihilation rate $\lambda$ which are large, and thus out of reach of a perturbative analysis performed around the reaction-less Gaussian fixed point. 

As for BARW-PC, the perturbative studies of \cite{cardy96} showed the existence of a new universality class different from DP, and of a new fixed point for $d$ smaller than a new critical dimension $d_c\simeq 4/3$. This behavior had already been predicted as a consequence of the additional symmetry \cite{grassberger84}, although some early studies \cite{grassberger89} were not conclusive with respect to this new universality class. In the NPRG context there have also been studies of BARW-PC \cite{canet3}, which seem to confirm the existence of a new fixed point for $d<d_c\simeq 4/3$. Within both methods the appearance of this new fixed point is associated with a change of stability of Renormalization Group  fixed point corresponding to Pure Annihilation (PA, a theory without branching reactions): for $d>d_c$, the branching $\sigma$ is a relevant perturbation and only an active phase exists, whereas for $d<d_c$, $\sigma$ is irrelevant and an absorbing phase with the properties of PA at long distances is 
expected for small $\sigma$.   

In this work (i) we obtain exact and closed equations for all response functions of the PA model and (ii) we show how to perform an expansion in $\sigma$ around this model. Since our approach is valid for any value of the annihilation rate $\lambda$, we obtain exact results at small $\sigma$. 

For BARW-DP we show that an active to absorbing phase transition exists in all dimensions $d$ and we compute the non-universal threshold values $\lambda_{th}(d)$ above which it occurs for two specific microscopic realizations of the system.

For BARW-PC we show, in disagreement with both the 2-loop perturbative and Local Potential Approximation (LPA) results, that the stability of the PA fixed point does not change between one and two dimensions. This contradicts the existing scenarios explaining the existence of a phase transition in $d=1$. We propose an alternative scenario that reconciles all existing results. It is based in the appearance in 
a dimension $d \in ]1,2]$ of two fixed points that split apart as $d$ is decreased. The fixed point with the smallest fixed point value for the branching rate should be fully attractive, while the other one should have a repulsive direction associated with the active to absorbing phase transition belonging to the PC universality class.

Some of these results have been recently presented in an abridged form in \cite{barwprl}.

The paper is organized as follows. In Section \ref{ft} we give a quick overview of the application of field theory to the study of out of equilibrium statistical systems. The interested reader is nonetheless strongly advised to read more general reviews of these methods \cite{henkel,tauber05,lee94}. In Section \ref{pa} we show a method to find all generalized response functions in the steady state of the simple reaction diffusion system corresponding to PA. In Sections \ref{odd} and \ref{even} we propose an expansion around this solution, in order to analyze BARWs in both universality classes, and to answer some specific questions concerning their phase diagrams. We have decided, in order to make the proofs simple, to use along the main part of the article derivations based on resummations of perturbative series. For completeness, however, we give non perturbative proofs (beyond an all-loop order analysis) of our results in the appendixes, as well as  presenting some other technical details.

\section{Field theory for BARW}\label{ft}

There are many known methods in the literature for the mapping of out of equilibrium problems onto field theories \cite{lee94,janssen76,doi76b}. In the case of reaction-diffusion processes, a field theory can be constructed in a standard way by using the Doi-Peliti formalism \cite{doi76b}, the idea of which is to re-express the Master Equation for the occupation probabilities in a lattice system using creation and annihilation operators in an abstract Fock space, followed by a coherent-state path integral representation, and (optionally) the use of a continuum limit for the lattice. As a result of this procedure, one obtains a functional integral (the so-called generating functional) 
\begin{equation}\label{ZJ}
{\cal Z}[J,\hat J] = \int {\cal D}\phi {\cal D} \hat \phi \exp \left(-S[\phi,\hat \phi]+\int_{x} J \phi+ \hat J \hat \phi \right)
\end{equation}
with an appropriate action $S[\phi,\hat \phi]$, which captures exactly the microscopic reactions. Here we have introduced the notation, to be used throughout
\begin{equation}
 x=(\mathbf{x},t) \qquad \mathrm{and}\qquad p=(\mathbf{p},\nu)
\end{equation}
where the last convention will be used in Fourier space. We also introduce some notation for the integrals
\begin{align}
 \int_x&=\int  d^dx\, d t & \int_p &= \int \frac{d^d p}{(2\pi)^d} \frac{d\omega}{2\pi}
\end{align}

The time-dependent statistical correlation and response functions can then be computed from ${\cal Z}$, by functional derivation w.r.t. to the sources $J$ and $\hat J$. In this context, the expected value of the field $\phi(x)$ is associated with the local density of $A$ particles, and the response field $\hat \phi(x)$ allows for the computation of response functions.

For general processes of the type $A\xrightarrow{\sigma_m} (m+1)A$ and  $kA\xrightarrow{\lambda_k} \emptyset$, with diffusion constant $D$, this procedure yields (ignoring initial conditions, which play no role in the long time stationary state, see for example \cite{tauber05})
\begin{multline}\label{actiongeneral}
 S[\phi,\hat \phi]  = \int_x
 \Big( \hat \phi(x)\,\big(\partial_t  - D \mathbf{\nabla}^2\big)\phi(x) - \lambda_k \big(1-\hat \phi(x)^k\big)\phi(x)^k\\
 +\sigma_m\,\big(1-\hat \phi(x)^m\big)\hat \phi(x)\phi(x) \Big)
\end{multline}

Diffusion is responsible for the kinetic part (corresponding to Brownian motion). Reactions give rise to interaction terms in the potential-like part of the action. A perturbative expansion can be set around the exactly solvable Gaussian part of the action in the usual way \cite{ZINN}. One can perform a perturbative expansion \cite{cardy96} to approximate the correlation and response functions of the theory at any desired order in $\lambda_k$, $\sigma_m$ and $\epsilon=d_c-d$, with $d_c$ the upper critical dimension, above which mean field results are expected to give a good description of the universal properties of the system. As stated in the introduction, for the purposes of the present work we will concentrate on reactions involving a minimal number of particles. These are enough to characterize the universal properties of these systems, and, for non-universal properties, they can be seen as the simplest examples.

The connected correlation and response functions of a theory will be written
\begin{multline}\label{correlation}
 G^{(n,m)}(x_1,\dots,x_n,\hat x_1,\dots, \hat x_{m})=  \\ \langle \phi(x_1)\dots \phi(x_n) \hat \phi(\hat x_{1})\dots \hat \phi(\hat x_{m})\rangle_c
\end{multline}
which are generated by taking derivatives of the logarithm of the generating functional $\log {\cal Z}[J,\hat J]$. These can be obtained in a perturbative series using connected Feynman diagrams. In this work we will mostly work with the vertex functions $\Gamma^{(n,m)}$, the amputated 1PI functions of the theory, which include all the information coming from fluctuations in the system. The generating functional $\Gamma[\langle \phi \rangle,\langle \hat \phi\rangle]$ for the $\Gamma^{(n,m)}$ vertices is given by the Legendre transform of the connected generating functional $\log {\cal Z}[J,\hat J]$.

It is often convenient to perform a shift in the fields, of the form \cite{cardy96}
\begin{equation}\label{shift}
 \hat \phi(x) = 1+ \bar \phi(x)
\end{equation}
which allows for some simplifications in the functional form of the interaction potential, and is needed to make the nexus between BARW-DP and Directed Percolation. This shift is not convenient in the BARW-PC case however, where it obscures the presence of the related parity conserving symmetry. 

In the case of out of equilibrium models, special care must be taken with respect to the causal structure of the theory. In this regard, actions such as (\ref{actiongeneral}), given by the Doi-Peliti formalism, implicitly require the use of the It\^o prescription, in which all quantities are evaluated with the convention that the Heaviside function $\Theta(t)$ is zero for $t=0$ \cite{vankampen,gardiner}. In perturbation theory it is relatively easy to implement the so-called It\^o prescription, as it amounts to force closed propagator loops to be zero \cite{cardy96,lee94}. Non-perturbative equivalent results are given in \cite{Delamotte:ito} and in Appendix \ref{causal}.

Diagrammatically, each $\bar\phi \phi$ propagator can be represented as a line with an arrow going from $\bar \phi$ to $\phi$, and each such propagator carries a Heaviside function of time, expressing causality. We use in the following the diagrammatic convention of drawing only $\bar \phi$-$\phi$ propagators (that is, the function $G(p)=\left[\Gamma^{(1,1)}(-p)\right]^{-1}$) but we include, if allowed
in a given model, insertions of $\Gamma^{(2,0)}$ or $\Gamma^{(0,2)}$ as vertices.

\section{Pure Annihilation}\label{pa}

In this Section we study the simplest case of a reaction-diffusion system, PA, in which the only reaction in the
system is annihilation by pairs of diffusing particles $A+A \to \emptyset$, with a probability rate $\lambda$. 
Later we will use the exact solution for this particular system as the starting point of a perturbative expansion, in order to study more general BARW at small branching rates.
It is easy to prove \cite{tauber05,peliti} that this system belongs to the same universality class as pure coagulation, in which the only reaction is $A + A \to A$. In the following we will use the PA model but the pure coagulation case can be analyzed in a similar way.

After implementing the Doi-Peliti procedure and performing a shift in the response fields, Eq. (\ref{shift}), the bare action $S^{PA}$ can be written \cite{cardy96,tauber05}
\begin{equation}
\label{S_PA}
S^{PA}[\bar \phi, \phi] = \int_x\Big(\bar \phi (\partial_t - D \mathbf{\nabla}^2) \phi + \lambda \bar \phi (\bar \phi +2)\phi^2 \Big). 
\end{equation}
As said before, we only analyze the steady state where all correlation functions are zero, since the system always approaches the empty state in the long time limit. However, even in this state, the response functions are non trivial, and are governed in the infrared (IR, that is to say, for momenta and frequencies smaller than the scale set by $\lambda$) by a non-trivial fixed point of the RG equations, for $d<2$. In the following we speak of ``correlation functions'' in a generalized sense, including response functions.

As it stands, this theory shows a certain resemblance with the standard $\phi^4$ scalar field theory. However, symmetry and causality properties allow for a greatly simplified analysis. We first show that for the PA model all $\Gamma^{(n,m)}$ functions can be obtained from the $\Gamma^{(n,n)}$, vertices with the same number of incoming and outgoing legs. This is quite clear perturbatively, but we give in the following a non-perturbative proof based on a Ward identity for a rescaling transformation. Secondly, we deduce a general identity yielding a closed equation for any $\Gamma^{(n,m)}$. It is easy to verify that the $\Gamma^{(1,1)}$, $\Gamma^{(2,1)}$ and $\Gamma^{(2,2)}$ vertices thus obtained coincide with the results of \cite{cardy96,peliti,wiese98,frey}. We show in Appendix \ref{gam33} how to compute $\Gamma^{(3,3)}$ from our method.

\subsection{Rescaling Ward identity}

Let us start by studying a generalization of PA with action $\tilde S^{PA}$, where couplings for the cubic and quartic terms are independent.
\begin{equation}\label{PAgeneralized}
\tilde S^{PA}[\bar \phi, \phi] = \int_x \Big(\bar \phi (\partial_t - D \mathbf{\nabla}^2) \phi + \lambda_3 \bar \phi \phi^2+ \lambda_4 \left(\bar \phi \phi\right)^2 \Big). 
\end{equation}
Let us consider the Ward identity \cite{ZINN} associated with the infinitesimal field transformation 
\begin{align}
 \phi (x) &\to (1+\epsilon)\phi(x) \notag \\
 \bar \phi (x) &\to (1-\epsilon)\bar \phi(x) \label{u1} 
\end{align}

When $\lambda_3 =0$ this is a symmetry of the action, but the cubic term breaks it explicitly. We can nevertheless obtain a Ward identity associated with this transformation by performing (\ref{u1}) as a change of variables in the expression for ${\cal Z}[J,\bar J]$, given in Eq. (\ref{ZJ}):
\begin{equation}\label{ward2}
 0=\epsilon\int_{x} \langle J \phi  -\bar J \bar \phi+\lambda_3 \phi^2 \bar\phi \rangle_{J,\bar J} 
\end{equation}
Here the mean value $\langle\dots  \rangle_{J,\bar J} $ is computed in the presence of the sources $J$ and $\bar J$. 
The term proportional to $\lambda_3$ can be written as a derivative w.r.t. $\lambda_3$ of the generating functional of connected correlation functions. By Legendre transforming Eq. (\ref{ward2}), one deduces the Ward identity
\begin{equation}\label{ward_u1}
 -\lambda_3\frac{\partial \Gamma}{\partial \lambda_3}+ \int_{x} \left( \phi \frac{\delta \Gamma}{\delta \phi} -\bar \phi \frac{\delta \Gamma}{\delta \bar \phi} \right)=0 
\end{equation}
This equation can be derived w.r.t. $\phi$ and $\bar \phi$ fields and evaluated at zero field, yielding
\begin{equation}
 (n-m)  \Gamma^{(n,m)}=\lambda_3 \frac{\partial \Gamma^{(n,m)}}{\partial \lambda_3}
\end{equation}
where $\Gamma^{(n,m)}$ is a function of $(x_1,\ldots,x_n,\bar x_1, \ldots, \bar x_m)$.

Since, perturbatively, $\Gamma^{(n,m)}$ can only involve positive powers of $\lambda_3$, this equation shows that $\Gamma^{(n,m)} \sim \mathcal{O}(\lambda_3^{n-m})$ when $n\geq m$, and that for PA $\Gamma^{(n,m)}$ contains exactly $(n-m)$ third-order bare vertices. We conclude that all $\Gamma^{(n,n)}$ vertices can be computed directly from the action with $\lambda_3=0$ and that
\begin{equation}\label{propPA}
 \Gamma^{(n,m)}(x_1,\ldots,x_n,\bar x_1, \ldots, \bar x_m)=0 \qquad \mathrm{if}\,n< m
\end{equation}
which simplifies the study of this system.

Given these results one can conclude that for any correlation function, the perturbative expansion in $\lambda_3$ is, being in fact a polynomial, exact at a {\it finite} order. In order to calculate the connected correlation function $G^{(n,m)}$ (with $n>m$), one can expand the functional integral at order $\lambda_3^{n-m}$:
\begin{widetext}
\begin{equation}
G^{(n,m)}(x_1,\dots,x_n,\bar x_1,\dots, \bar x_{m})=  \lambda_3^{n-m}\left.\langle \phi(x_1)\dots \phi(x_n) \bar \phi(\bar x_{1})\dots \bar \phi(\bar x_{m}) \times \Big(\int_{x}  \bar \phi \phi^2\Big)^{n-m}\rangle_c\right|_{J=\bar J=0,\lambda_3=0}
\end{equation}
\end{widetext}
(using the unique decomposition of $G^{(n,m)}$ in terms of 1PI vertices \cite{ZINN}) reducing its calculation to the knowledge of correlation functions of the $\lambda_3=0$ model (which only contains $\Gamma^{(n,n)}$ vertices). This shows that the building blocks of the PA model are the vertex functions with an equal number of incoming and outgoing legs that can be calculated at $\lambda_3=0$.

\subsection{An identity for the $\Gamma^{(n,m)}$ vertices}

\begin{figure}[ht]
\includegraphics[width=0.45\textwidth]{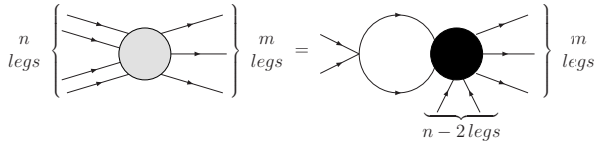}
\caption{Generic form of a diagram contributing to $\Gamma^{(n,m)}$ and that involves at least one loop in PA. Left hand side: diagrammatic representation of a generic $\Gamma^{(n,m)}$ vertex. Right hand side: general structure for such vertices in PA, the black blob is a connected and amputated Green function that has to comply with some requisites, see text.}
\label{gamnm}
\end{figure}

We now present an identity allowing us to obtain a closed equation for any $\Gamma^{(n,m)}$. It can be most conveniently written at the diagrammatic level: any diagram contributing to $\Gamma^{(n,m)}$ which includes at least one loop has the structure shown in Fig. \ref{gamnm} (that is: any 1PI perturbative diagram begins with a 4-legs bare vertex).
The black blob denotes a sub-diagram that is constrained by the condition that the full diagram must be 1PI. In particular, it means that this sub-diagram must be connected (and with amputated external legs). Now, any connected diagram with $n$ incoming and $m$ outgoing legs has a unique tree decomposition in terms of 1PI sub-diagrams having at most these numbers of legs. By summing all possible diagrams and permutations compatible with the 1PI structure of the full diagram, we obtain a closed equation that relates any $\Gamma^{(n,m)}$ with vertices $\Gamma^{(s,l)}$ with a lower number of legs. A non-perturbative proof (not based on an all-order analysis) of this general property is given in Appendix \ref{SD}, by using NPRG techniques (to which we give an introduction in Appendix \ref{nprg}). 

Notice that, as explained in the previous Section, $\Gamma^{(n,n)}$ vertices can be calculated at $\lambda_3=0$. Now, for  $\lambda_3=0$ the $U(1)$ transformation (\ref{u1}) is a symmetry of the action, and the fields $\phi$ and $\bar \phi$ play a symmetric role. Accordingly, this same construction can be performed singularizing two outgoing legs in the case of $\Gamma^{(n,n)}$ vertices. 

\begin{center}
\begin{figure}[ht]
\includegraphics[width=0.45\textwidth]{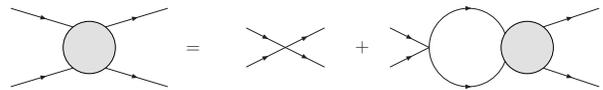}
\caption{Closed equation for $\Gamma^{(2,2)}$ in PA.}
\label{gam22}
\end{figure}
\end{center}

In order to be concrete, let us analyze the identity given in Fig. \ref{gamnm} for the simplest vertices. For $\Gamma^{(1,1)}$ this gives a well-known non-renormalization property: there is no correction to $\Gamma^{(1,1)}$ in PA. This is due to the fact that no diagram such as the one presented in Fig. \ref{gamnm} can be drawn with a single incoming leg. This no-field-renormalization condition implies that the critical exponents $\eta$ and $z$ have their mean field values, $\eta=0$ and $z=2$. Concerning $\Gamma^{(2,2)}$, the result is less trivial. Given that there are only two incoming legs and that in a theory without cubic vertices all connected diagrams with four external legs are 1PI, one arrives at
the closed equation (see Fig. \ref{gam22})
which reads:
\begin{multline}
\label{eqGamma22}
\Gamma^{(2,2)}(p_1,p_2,\bar p_1,\bar p_2)= 4 \lambda_4 - 2 \lambda_4 \int_q G(q) \\ 
\times G(p_1+p_2-q) \Gamma^{(2,2)}(q,p_1+p_2-q,\bar p_1,\bar p_2)
\end{multline}
whose solution is of the form (see Appendix \ref{nprgvertex} for a proof)
\begin{equation}\label{deflambda}
\Gamma^{(2,2)}(p_1,p_2,\bar p_1,\bar p_2)= 4 l_4(p_1+p_2)
\end{equation}
By substituting (\ref{eqGamma22}) in (\ref{deflambda}) we find
\begin{equation}
\label{lambda(p)}
l_4(p)= \frac{\lambda_4}{1+ 2 \lambda_4 \int_q G(q) G(p-q)}
\end{equation}

\begin{center}
\begin{figure}[ht]
\includegraphics[width=0.45\textwidth]{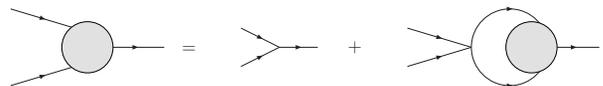}
\caption{Diagrammatic identity for $\Gamma^{(2,1)}$ in PA.}
\label{gam21}
\end{figure}
\end{center}

For $\Gamma^{(2,1)}$ the identity in Fig. \ref{gamnm} becomes that of Fig. \ref{gam21}, which can be written as
\begin{multline}
\label{eqGamma21}
\Gamma^{(2,1)}(p_1,p_2,\bar p)= 2 \lambda_3 - 2\lambda_4 \int_q G(q) \\ \times G(p_1+p_2-q) \Gamma^{(2,1)}(p_1+p_2-q,q,\bar p)
\end{multline}
We can show (see Appendix \ref{nprgvertex}) that this implies that $\Gamma^{(2,1)}(p_1,p_2,\bar p)$ depends only on $\bar p$. We thus define
\begin{equation}\label{deflambda3}
 \Gamma^{(2,1)}(p_1,p_2,\bar p)= 2 l_3(\bar p)
\end{equation}
By substituting Eq. (\ref{deflambda3}) into Eq. (\ref{eqGamma21}) we find
\begin{equation}
l_3(\bar p)= \frac{ \lambda_3}{1+ 2 \lambda_4 \int_q G(q) G(\bar p-q)}
\end{equation}
Dividing $\Gamma^{(2,1)}$ by $\Gamma^{(2,2)}$ one observes that their quotient is equal to $\lambda_3/(2 \lambda_4)$, so that the relation between three and four point vertices is not renormalized. Of course, for the actual PA model one must take $\lambda_3=2 \lambda_4=2\lambda$, and conversely $l(p)=l_4(p)$. In the rest of the manuscript, we only consider this case unless otherwise stated.
 
In Appendix \ref{nprgvertex} it is shown, by using NPRG equations, that these expressions are in fact non-perturbative (they are valid beyond an all-order perturbative analysis). These expressions have already been obtained before for the vertices with two incoming legs \cite{cardy96,peliti,wiese98}, as a sum over bubbles. The interesting point is that the present analysis applies to any $\Gamma^{(n,m)}$ vertex in PA. As an example, in appendix \ref{gam33} the equation for $\Gamma^{(3,3)}$ is obtained. Unfortunately, for $n>2$ the corresponding equations must be solved numerically.

Now that we have a method to calculate all correlation functions in PA, we can study BARW by means of a perturbative expansion in the branching rate $\sigma$. We stress that a perturbative expansion on a coupling constant around a non-Gaussian model, such as PA, is a priori difficult to perform. 

To end this section, notice that all the results above are independent of the space dimension $d$. This allows us to make predictions independently of the upper critical dimension $d_c$ of the BARW systems studied below.

\section{BARW - DP}\label{odd}

In this Section we consider the simplest BARW-DP model, where the only reactions are $A\to 2A$ and $2A\to \emptyset$. More general cases in the DP universality class can be considered as well using the same methods. The microscopic action for this model reads, after the shift in the response fields (see Eq. (\ref{actiongeneral}))
\begin{equation}
S^{DP}= \int_x \Big(\bar \phi (\partial_t - D \mathbf{\nabla}^2) \phi + \lambda \bar \phi (\bar \phi +2)\phi^2 - \sigma \bar \phi (\bar \phi + 1)\phi \Big) 
\end{equation}
We now show how to perform a systematic expansion in $\sigma$ while keeping a finite $\lambda$. This expansion is particularly well suited for properties of the model that take place at small $\sigma$, but at values of $\lambda$ that can be out of reach of a perturbative expansion around the Gaussian theory. As mentioned in the introduction, the transition between the active and the absorbing phases in this model takes place, for $d>2$, at values of $\lambda$ larger than a threshold $\lambda_{th}$, which make the calculation of the phase diagram impossible within the usual perturbative analysis in these dimensions. As this threshold corresponds to $\sigma$ arbitrarily small, the value or $\lambda_{th}$ is computable in an exact way at the leading order of the expansion in $\sigma$ that we detail in the following. We stress, however, that the calculation of this quantity is just a specific example of an application of the expansion in $\sigma$, which may be used for more general purposes.

In order to analyze BARW-DP it is useful to consider, as in PA, a generalization of $S^{DP}$ with independent couplings. We then consider the action
\begin{equation}
\tilde S^{DP}= \int_x \Big(\bar \phi (\partial_t - D \mathbf{\nabla}^2) \phi + \lambda_3 \bar \phi \phi^2 + \lambda_4 (\bar \phi\phi)^2 - \sigma_2 \bar \phi \phi -\sigma_3 \bar \phi^2 \phi\Big) 
\end{equation}

As in the case of PA, one can deduce a Ward identity for the rescaling transformations (\ref{u1}), which in this case reads
\begin{equation}\label{ward_u1-odd}
 -\lambda_3\frac{\partial \Gamma}{\partial \lambda_3} +\sigma_3\frac{\partial \Gamma}{\partial \sigma_3} +\int_{x} \left( \phi \frac{\delta \Gamma}{\delta \phi} -\bar \phi \frac{\delta \Gamma}{\delta \bar \phi} \right)=0 
\end{equation}
that leads us to
\begin{equation}\label{wardodd2}
 (n-m)  \Gamma^{(n,m)}=\lambda_3 \frac{\partial \Gamma^{(n,m)}}{\partial \lambda_3} - \sigma_3 \frac{\partial \Gamma^{(n,m)}}{\partial \sigma_3}
\end{equation}
where $\Gamma^{(n,m)}$ is a function of $(x_1,\ldots,x_n,\bar x_1, \ldots, \bar x_m)$. The solution of (\ref{wardodd2}) implies the following relation for $\Gamma^{(n,m)}$
\begin{equation}
\label{propodd}
\Gamma^{(n,m)}(\sigma_2,\sigma_3,\lambda_3,\lambda_4) = \sigma_3^{m-n} \gamma^{(n,m)}(\sigma_2,\sigma_3 \lambda_3,\lambda_4)
\end{equation}
for $m>n$, with $\gamma^{(n,m)}$ a regular function of its arguments (in particular for $\sigma_3=0$). This is nothing but the well known result of perturbation theory, which states, putting aside a re-scaling of vertices, that cubic couplings appear only via their product. At leading order in $\sigma_3$, Eq. (\ref{propodd}) shows that
\begin{equation}\label{sigmaexpodd}
 \Gamma^{(n,m)}\sim\mathcal{O}(\sigma_3^{m-n})\qquad \mathrm{for} \,\, n<m
\end{equation}
and that the calculation at leading order can be done at $\lambda_3=0$.

In order to perform the $\sigma_3$-expansion one can consider the generating functional (\ref{ZJ}) and expand the exponential term
\begin{multline}
 \mathcal Z=\int \mathcal{D} \phi \mathcal{D} \bar \phi \exp \Big( -\tilde S^{DP}\big|_{\sigma_3=0} +\int_x J \phi+ \bar J \bar \phi  \Big) \\ \times\sum_{k=0}^\infty \frac{1}{k!} \left( \sigma_3 \int_x \bar \phi^2 \phi  \right)^k
\end{multline}
In this way, the calculation to any order in $\sigma_3$ of any correlation function is reduced to the calculation of higher order correlation functions in a modified PA that includes a mass-like $\sigma_2$ term. It is worth mentioning that the methods presented in the previous Section work as well in the model including a $\sigma_2$ term. When and if this $\sigma_2$ term is not necessary to make the theory IR safe it is possible to expand in $\sigma_2$ as well as $\sigma_3$ and this is what we are going to do in practice.

As a final comment with respect to the $\sigma$-expansion, it is important to notice that it generates a \emph{convergent} series, something not very common when dealing with perturbative expansions in field theories. This property follows from Lebesgue's dominated convergence theorem \cite{folland}, given that we have under nonperturbative control the PA model (as  shown in Appendixes \ref{SD}, \ref{causal} and \ref{nprgvertex}).

This convergence property can be most easily seen by working with a zero dimensional toy model
\begin{equation}
 Z=\int dx\, e^{-\lambda x^4+\sigma x^3}
\end{equation}
Defining
\begin{equation}
 f_n(x) =\sum_i^n e^{-\lambda x^4} \frac{1}{n!} \sigma^n x^{3n}
\end{equation}
we see that the integrands
\begin{equation}
 0\leq |f_n(x)| \leq A e^{-\lambda' x^4/2}  
\end{equation}
for some large enough constant $A$ and appropriate $\lambda' > \lambda$. From the dominated convergence theorem we know that the sequence of integrals of functions $f_n(x)$ converge to the integral of the function $f(x)=\lim_{n\to \infty} f_n(x)$. A similar reasoning applies in the case of the $(d+1)$-dimensional model, at least for the model defined on a lattice and in a finite volume.

\subsection{Threshold of the active-to-absorbing transition for BARW-DP}

Let us consider as a specific example the calculation of the threshold $\lambda_{th}$ for the existence of an active-to-absorbing phase transition in BARW-DP. Notice that this threshold value is non-universal, as would be a critical temperature in an equilibrium model. The question of whether a phase transition in this system is continuous or discontinuous can not be addressed within the $\sigma$-expansion, because the dependency on an external background field should be taken into account, and in this work we are considering PA at vanishing external field. However, a phase transition of the continuous type is a priori known to exist in these systems, following Monte Carlo results \cite{canet2}. Enforcing this, a second order phase transition has been rigorously proven to take place in a related BARW system, known as the contact process \cite{refdick}. 

In order to check for the presence of such a continuous phase transition in BARW-DP, it is enough to study the behavior of $\Delta=\Gamma^{(1,1)}(p=0)$ as a function of the annihilation rate $\lambda$. In fact, we can detect this phase transition by looking for the zeros of $\Delta$, which correspond to a divergence in the 
correlation length \cite{Delamotte:2007pf}.

\begin{center}
\begin{figure}[ht]
\includegraphics[width=0.45\textwidth]{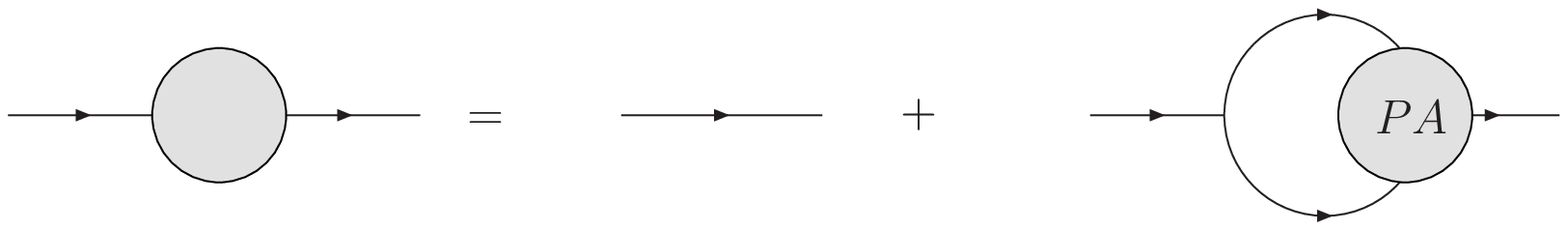}
\caption{Closed equation for $\Gamma^{(1,1)}$ at first order in $\sigma$ in BARW-DP.}
\label{gam11odd}
\end{figure}
\end{center}

Given that $\lambda_{th}$ corresponds to the transition value of $\lambda$ when $\sigma\to 0^+$, an analysis at leading order in $\sigma$ allows for an exact calculation of $\lambda_{th}$. Following the lines of the previous discussion, an equation for $\Gamma^{(1,1)}(p)$ at order $\mathcal{O}(\sigma)$ can be represented in the diagrammatic form of Fig. \ref{gam11odd}, that can be written  
\begin{align}
\Gamma^{(1,1)}(p)& =-\sigma \notag \\
& +\sigma \int_{q} G(q)G(p-q) \Gamma^{(2,1)}(q,p-q,-p) + \mathcal{O}(\sigma^2)\notag \\
&= -\sigma +4\sigma l(p) \int_{q} G(q)G(p-q)  + \mathcal{O}(\sigma^2)\label{flow_delta}
\end{align}
In the last line of (\ref{flow_delta}) we have evaluated the propagator $G(p)$ and the vertex $\Gamma^{(2,1)}(q,p-q,-p)$ at order zero in $\sigma$, and consequently replaced this last function by $4 l(p)$ (see Eq. (\ref{lambda(p)}), remember that we consider $l(p)=l_4(p)$). 

\begin{center}
\begin{figure}[ht]
\includegraphics[width=0.45\textwidth]{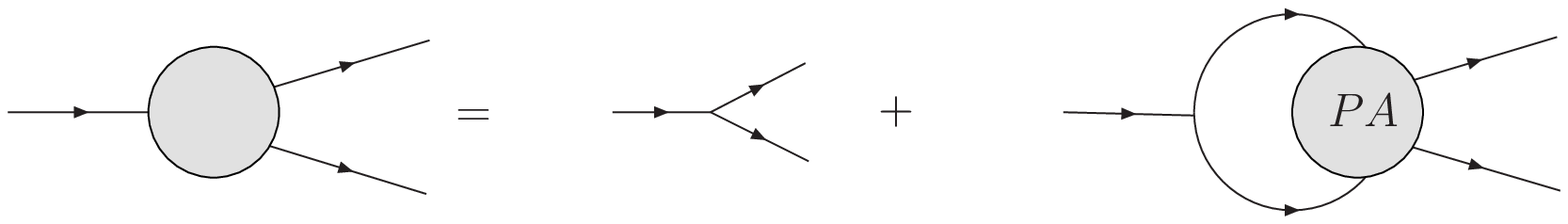}
\caption{Closed equation for $\Gamma^{(1,2)}$ at first order in $\sigma$ in BARW-DP.}
\label{gam12odd}
\end{figure}
\end{center}
As a side note, observe that we could have just as well written an equivalent equation for $\Gamma^{(1,2)}$ (see Fig. \ref{gam12odd}), which reads at order $\sigma$ 
\begin{align}\label{flow_delta2}
\Gamma^{(1,2)} (p_1,\bar p_1,\bar p_2) & =-2\sigma
+2\sigma \int_q  G(q) G(p_1-q) \notag \\ &\quad \times \Gamma^{(2,2)}(p_1-q,q,\bar p_1,\bar p_2)  + \mathcal{O}(\sigma^2)\notag \\
&= -2\sigma +8\sigma \int_q  G(q) G(p_1-q) l(p_1)  + \mathcal{O}(\sigma^2)
\end{align}
As before, the $\Gamma^{(2,2)}(q,p_1-q,\bar p_1,\bar p_2)$ vertex can be taken at order $\sigma^0$, that is, it can be taken to be equal to $4l(p_1)$. Expressions (\ref{flow_delta}) and (\ref{flow_delta2}) imply that
\begin{equation}
\Gamma^{(1,2)}(p_1,\bar p_2,\bar p_3)=2\Gamma^{(1,1)} (p_1)  + \mathcal{O}(\sigma^2)
\end{equation}
which states that at first order in $\sigma$, the bare relation between the $(1,1)$ and $(1,2)$ vertices is maintained.

Returning to our problem, we can look for a second order phase transition by studying the behavior of $\Delta$.  One needs the non-universal
value $l(p=0)$ that can be obtained by evaluating Eq. (\ref{lambda(p)}) at $p=0$:
\begin{equation}
 l(p=0) = \frac{\lambda}{1+2 \lambda I(d)}
\end{equation}
where
\begin{equation}
I(d)=\int_q G(q)G(-q) 
\end{equation}
By substituting the expression for $l(p=0)$, and evaluating (\ref{flow_delta}) at $p=0$ one arrives at
\begin{equation}
 \Delta = -\sigma +4 \sigma  \frac{\lambda I(d)}{1+2\lambda I(d)} + \mathcal{O}(\sigma^2)
\end{equation}
which for $\Delta=0$ implies a threshold value
\begin{equation}
\label{lambdac}
 \lambda_{th}=\frac{1}{2 I(d)}
\end{equation}
To evaluate $\lambda_{th}$, we need to take into account that the properties of a phase diagram are not universal and depend on the specific form of the theory at small distances. This is as in equilibrium statistical mechanics, where critical temperatures depend on the specific form of the lattice. We will consider two particular microscopic forms for the model. The first one corresponds to the model defined on a hyper-cubic lattice with lattice spacing $a$. The second corresponds to a `continuum' version where a UV cut-off is imposed at a finite (but large) scale $\Lambda$.

For the hyper-cubic lattice, the propagator reads 
\begin{equation}
 G(q)=\frac{1}{i\omega+\frac{2D}{a^2}\sum_{i=1}^d (1-\cos(a q_i))},
\end{equation}
and the integral in (\ref{lambdac}) becomes
\begin{align}
 I(d)&=\int \frac{d\omega}{2\pi}\int_{-\pi/a<q_i\leq\pi/a}\frac{d^dq}{(2\pi)^d}G(q)G(-q)\nonumber\\
&=\frac 1 2 \int_{-\pi/a<q_i\leq\pi/a}\frac{d^dq}{(2\pi)^d}\frac{1}{\frac{2D}{a^2}\sum_{i=1}^d (1-\cos(a q_i))}\nonumber\\
&=\frac{a^{2-d}} {4\,D}  \int_{-\pi<q_i\leq\pi}\frac{d^dq}{(2\pi)^d}\frac{1}{\sum_{i=1}^d (1-\cos(q_i))}
\end{align}
where the integral over $\omega$ has been performed by using the residues' theorem. The remaining integral must be calculated numerically. In Table \ref{table}, the value of the resulting threshold coupling is given. Previous results from Monte-Carlo simulations and approximated NPRG equations \cite{canet2,odor1} are in good agreement with these exact ones. This same general structure of the phase diagram has also been shown to exist in other models in the DP universality class \cite{odor2}.

An interesting property observed in \cite{canet2} is that
$\lambda_{th}$ seems to grow linearly with $d$. In \cite{canet4}, a single-site approximation scheme that is argued to become exact in the large $d$ limit in a hyper-cubic lattice
was analyzed, and this linear behavior was obtained. In order to analyze such a behavior here, it is necessary to find the large-$d$ limit for the integral $I(d)$. For this purpose it is useful to re-write it in the following form:
\begin{equation}
 I(d) D a^{d-2}= \frac{1} {4\,d} \int_{-\pi<q_i\leq\pi}\frac{d^dq}{(2\pi)^d}\frac{1}{1-\big(\sum_{i=1}^d \cos(q_i)\big)/d}
\end{equation}
To solve it, one can imagine the various $\cos(q_i)$ as random variables with zero mean. By the strong law of large numbers, their mean $\big(\sum_{i=1}^d \cos(q_i)\big)/d$ tends to zero, except in a zero measure set. We are then tempted to substitute the limit inside the integrand and
obtain
\begin{equation}
 I(d) D a^{d-2}\stackrel{d\to\infty}{\sim} \frac{1} {4\,d}.
\end{equation}
This step is non trivial from a rigorous mathematical point of view, but turns out to be correct by using elaborate methods of real analysis \cite{note1}. As a consequence,
\begin{equation}
\lambda_{th}/D a^{2-d}\stackrel{d\to\infty}{\sim} 2 d
\end{equation}
in agreement with previous results \cite{canet4}.
\begin{center}
\begin{table}[tp]
\begin{tabular}{lcccc}
\hline
  $d$                   &   3  &  4   &  5   &   6   \\ \hline
 $\lambda_{th}/D a^{d-2}$ (this work)  & 3.96 & 6.45 & 8.65 & 10.7 \\
 $\lambda_{th}/D a^{d-2}$ (Monte-Carlo) \cite{canet2}  & 3.99  & 6.48 & 8.6 & 10.8 \\
\hline
\end{tabular}
\caption{\label{table} Values of the threshold coupling $\lambda_{th}$ for various dimensions $d$. Comparison of present exact results with Monte-Carlo \cite{canet2}.}
\end{table}
\end{center}

It is interesting to observe that expression (\ref{lambdac}) only depends on quantities that are calculated exactly in the Local Potential Approximation (LPA) of the NPRG, which is the lowest order of the Derivative Expansion. Only vertices at zero momenta are used and their exact equation turns out to be the same as the one that comes from the LPA (see Appendix \ref{nprg}). This a posteriori explains the success of the LPA in reconstructing the phase diagram of this model \cite{canet2}. 

However, as mentioned before, the phase diagram is a non-universal property that depends on the precise definition of the model in the ultraviolet. In particular, the value of the integral $I(d)$ is different if calculated in a discrete lattice or in the continuum with a given ultraviolet regularization. In the previous study done within the NPRG \cite{canet2}, a continuum version of the model was implemented, but the initial bare condition was imposed at a finite (but large) value of the microscopic scale $\Lambda$, which serves as a UV cut-off. In order to be able to compare our continuum non-universal results with those obtained in \cite{canet2} we will choose an UV regularization compatible with the NPRG procedure, as described in Appendix \ref{nprg}.

In the continuum regularized case, the integral to be calculated in order to make a direct comparison with the NPRG (see Appendix \ref{nprgvertex}) is (with the tilde indicating this second regularization)
\begin{align}
 \tilde I(d)&=\frac{1}{(4\pi)^{d/2}\Gamma(d/2)\,D} \int_0^\Lambda dq \,q^{d-1}\Big(\frac{1}{q^2}-\frac{1}{\Lambda^2}\Big)\nonumber\\
&=\frac{\Lambda^{d-2}}{(4\pi)^{d/2}\Gamma(d/2)\,D} \frac{2}{d(d-2)}.
\end{align}
This yields for this particular regularization
\begin{equation}
\tilde \lambda_{th}=\frac{\Lambda^{2-d}(4\pi)^{d/2}\Gamma(d/2)\,D d (d-2)}{4}
\end{equation}
Given that this integral is calculated in a closed form by using exclusively quantities evaluated at momentum $p=0$, we can check (see Appendix \ref{nprgvertex}) that it coincides exactly with the LPA equation for this same quantity. Indeed, our result recovers the numerical LPA solution of the NPRG of \cite{canet2} within a nine digit accuracy.

We can also compare the results coming from both lattice and continuum regularizations, as has been done in \cite{canet2,canet06}. As is explained there, one cannot do such comparison without fixing the relation between $\Lambda$ and the lattice spacing $a$. In \cite{canet2}, this
relation was fixed by multiplying the continuum results by $\exp(c (2-d))$ and fitting the constant number $c$, finding a very reasonable agreement up to $d=7$. However, we observe in the present results that the agreement is lost in higher dimensions where the continuum version
leads to
\begin{equation}
\tilde \lambda_{th}\Lambda^{d-2}/D\stackrel{d\to\infty}{\sim}\frac{(2\pi\,d/e)^{d/2}\sqrt{\pi}\, d^{5/2}}{4}.
\end{equation}
This indicates that the agreement between both results is only valid for a limited range of dimensions. In order to relate the results in a larger range of dimensions, one must consider a $d$-dependent relation between $\Lambda$ and $a$ or, as done here, take into account the precise ultraviolet regularization considered.

Finally, it is convenient to point out that for $d\leq 2$ an IR divergence of the integral in (\ref{lambdac}) takes place. This makes $\lambda_{th}=0$ in those dimensions, in agreement with the results of \cite{cardy96}. For this reason, for $d\leq 2$ it is not useful to expand the model at small $\sigma$ for a finite $\lambda$ in order to study the phase transition. Moreover, this also shows that in those dimensions the transition is dominated by IR effects, and correspondingly most of the dependence on the microscopic behaviour of the model is absent.

\section{BARW - PC}
\label{even}

Let us now consider BARW-PC, corresponding to the Parity Conserving/Genteralized Voter universality class. In this case, it is convenient not to shift the response field in order to make explicit the $\phi \to -\phi$, $\hat \phi \to -\hat \phi$ symmetry associated with conservation of the parity of the number of particles. The microscopic action for the BARW-PC model reads (see Eq. (\ref{actiongeneral}))
\begin{equation}\label{S-EVEN}
 S^{PC}[\phi, \hat \phi] = \int_x \Big(\hat \phi (\partial_t - D \mathbf{\nabla}^2) \phi + \lambda (\hat \phi^2 -1)\phi^2 \\ + \sigma (1-\hat \phi^2) \phi \hat \phi \Big)
\end{equation}
where the last term corresponds to the branching reaction $A \to 3A$ with rate $\sigma$.

The case $\sigma=0$ corresponds to Pure Annihilation, now written in terms of the non-shifted $\hat \phi$ field. This version of Pure Annihilation can again be solved following the same ideas as previously. Here, as opposed to the shifted case, we have the additional constraint that $\Gamma^{(n,m)}=0$ if $(n+m)$ is odd. 

Let us now show that in this version of PA
\begin{equation}\label{relPAnonshift}
\Gamma^{(n,m)}\sim \mathcal{O}(\lambda^{(n-m)/2}) \quad \mathrm{for}\, n\geq m 
\end{equation}
and zero otherwise. We again define a generalized action $\tilde S^{PC}$ with independent $\lambda_2$ and $\lambda_4$ couplings as in Eq. (\ref{PAgeneralized})
\begin{multline}\label{Smodpc}
 \tilde S^{PC}[\phi, \hat \phi] = \int_x \Big(\hat \phi (\partial_t - D \mathbf{\nabla}^2) \phi - \lambda_2 \phi^2 \\ +\lambda_4 \hat \phi^2 \phi^2  +\sigma_2 \hat\phi  \phi  - \sigma_4 \hat \phi^3 \phi \Big)
\end{multline}
First we set $\sigma_2$ and $\sigma_3$ equal to zero, in order to be in PA, and exploit the Ward identity for the infinitesimal transformation
\begin{align}
 \phi (x) &\to (1+\epsilon)\phi(x) \notag \\
 \hat \phi (x) &\to (1-\epsilon)\hat \phi(x) \label{u1-even} 
\end{align}
The argument is completely analogous to the one shown in Section \ref{pa}, yielding
\begin{equation}
 (n-m)  \Gamma^{(n,m)}=2\lambda_2 \frac{\partial \Gamma^{(n,m)}}{\partial \lambda_2}
\end{equation}
(with, as before, $\Gamma^{(n,m)}$ a function of $(x_1,\ldots,x_n,\bar x_1, \ldots, \bar x_m)$) from which Eq. (\ref{relPAnonshift}) follows.

It is easy to check that the equation for $\Gamma^{(2,2)}$ remains the same as in the shifted case, Eq. (\ref{eqGamma22}), and we thus define the function $l(p)$ again by means of Eq. (\ref{lambda(p)}). The vertex $\Gamma^{(2,0)}$ can be studied by following similar lines, and is found to be related to $l(p)$, by $\Gamma^{(2,0)}(p)=-2l(p)$. Also as before, $\Gamma^{(1,1)}$ is easily proven not to be renormalized in this version of PA.

Since we are interested in studying the $\sigma$-expansion around PA, it is useful to establish the equivalent of Eq. (\ref{sigmaexpodd}) regarding the order in $\sigma$ of the $\Gamma^{(n,m)}$. We again work with generalized couplings $\sigma_2$ and $\sigma_4$, using the modified action (\ref{Smodpc}) and we arrive, by using the rescaling Ward identity deduced from (\ref{u1-even}) (an identity similar to Eq. (\ref{ward_u1-odd})) at the relationship 
\begin{equation}
\label{propeven}
\Gamma^{(n,m)}(\sigma_2,\sigma_4,\lambda_2,\lambda_4) = \sigma_4^{(m-n)/2} \gamma^{(n,m)}(\sigma_2,\sigma_4 \lambda_2,\lambda_4) 
\end{equation}
for $m>n$, with $\gamma^{(n,m)}$ a regular function of its arguments (in particular for $\sigma_4=0$). This implies that $\Gamma^{(n,m)}\sim \mathcal{O}(\sigma^{(m-n)/2})$ if $m>n$. The details of the calculations leading to this property are completely analogous to those of Section \ref{odd} in BARW-DP.

\subsection{The stability of the PA fixed point}

\begin{center}
\begin{figure}[ht]
\includegraphics[angle=0,width=0.45\textwidth]{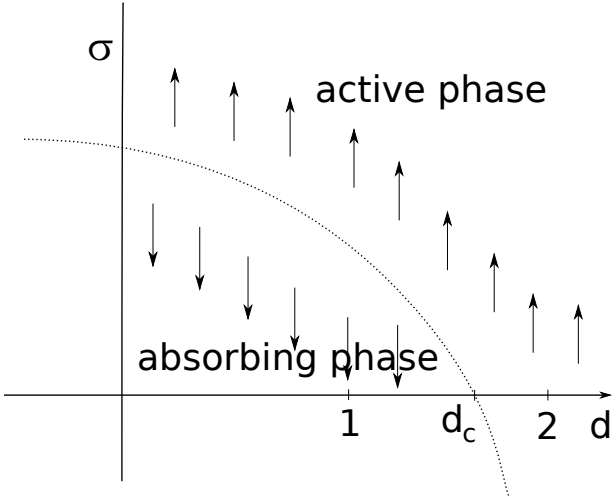}
\caption{Sketch of the relevance of the $\sigma$ perturbation in BARW-PC around the PA fixed point, as expected from \cite{cardy96,canet3}. The arrows show the direction of the RG flow for the coupling $\sigma$. Above $d_c$, $\sigma$ is relevant, whereas it is irrelevant below $d_c$. The dashed line represents the location of the fixed point $F^{PC}$ that crosses the PA fixed point at $d_c$ and that is associated with a phase transition below $d_c$.}
\label{graph1}
\end{figure}
\end{center}

One striking feature of the PC model is the existence of an active-to-absorbing phase transition in $d=1$. This is believed to be related to a change of stability of the PA fixed point in a dimension $d_c$ between one and two. Perturbatively, and also within the LPA, this change of stability occurs in the following way (see a schematic representation of this scenario in Fig. \ref{graph1} \cite{cardy96,canet3}). On one hand, in $d=2$, the Gaussian and PA fixed points merge so that, for dimensions close to two, the relevance of the branching reaction $A \xrightarrow{\sigma} 3A$ can be proven by canonical power counting arguments. On the other hand, at 1- and 2-loop orders an (upper) critical dimension $d_c>1$ is found such that for $d<d_c$ the coupling $\sigma$ becomes irrelevant around the PA fixed point which therefore becomes fully attractive. This change of stability occurs because a new fixed point, $F^{PC}$, crosses the PA fixed point at $d_c$ and in this dimension they both change their stability. Below 
$d_c$, this new fixed point is in the physically relevant quadrant $\lambda \geq 0$, $\sigma \geq 0$, has one unstable direction, and  is thus associated with the phase transition. The PA fixed point is then fully attractive for $d<d_c$ and describes the absorbing phase. Notice that the value of $d_c$ changes significantly between 1-loop -- where $d_c=4/3$ -- and 2-loops where $d_c\simeq 1.1$ ($d_c=4/3$ within the LPA). 

Some of these facts seem to be confirmed by other methods. In $d=1$ Monte-Carlo simulations of this model show indeed a new universality class \cite{odorrev,jensen}, and an exactly solvable model expected to be in the same universality class as BARW-PC shows a negative scaling dimension for $\sigma$: $d_\sigma=-1$ \cite{takayasu92}. This result ($d_\sigma=-1$) is identical to the prediction at order $\epsilon$ \cite{cardy96}. At two loop order, though, this value for $d_\sigma$ changes and gets smaller in magnitude, $d_\sigma \simeq -0.137$ at $d=1$ \cite{cardy96}. As will be argued in the following, this significative difference between MC and 2-loop results can be seen as a first indication that the results of \cite{takayasu92} are not entirely valid for this system.

\begin{center}
\begin{figure}[ht]
\includegraphics[width=0.45\textwidth]{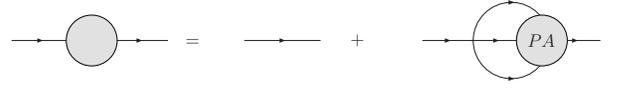}
\caption{Closed equation for $\Gamma^{(1,1)}$ in BARW-PC, at first order in $\sigma$.}
\label{gam11even}
\end{figure}
\end{center}

We now reanalyze the stability of the PA fixed point in the presence of the PC creation reaction, $A \xrightarrow{\sigma} 3A$, that we can determine exactly since our analysis is exact at small $\sigma$. The relevance of this coupling can be obtained from the flow of either 
$\Gamma^{(1,1)}$ or $\Gamma^{(1,3)}$, since both these functions are of order $\sigma$. However, the RG flow of  $\Gamma^{(1,1)}$ depends on the somewhat difficult to study $\Gamma^{(3,1)}$ vertex of PA (see Fig. \ref{gam11even}), and we prefer to study $\Gamma^{(1,3)}$.
\begin{center}
\begin{figure}[ht]
\includegraphics[width=0.45\textwidth]{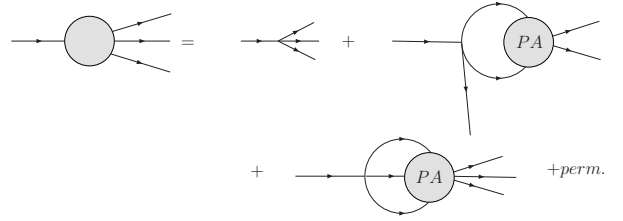}
\caption{Closed equation for $\Gamma^{(1,3)}$in BARW-PC at first order in $\sigma$.}
\label{gam13even1}
\end{figure}
\end{center}

\begin{center}
\begin{figure}[ht]
\includegraphics[width=0.45\textwidth]{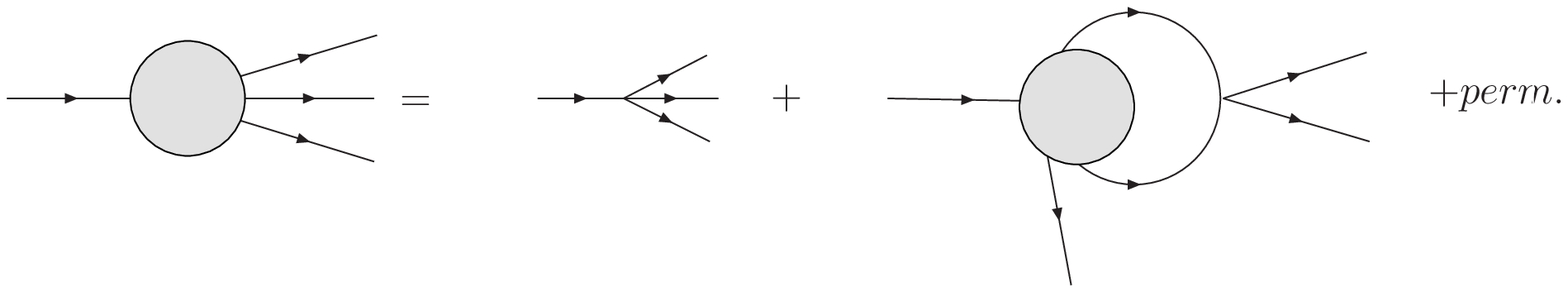}
\caption{Another possible closed equation for $\Gamma^{(1,3)}$ in BARW-PC, at first order in $\sigma$.}
\label{gam13even}
\end{figure}
\end{center}

At first order in $\sigma$, any diagram for $\Gamma^{(1,3)}$ will be of the form shown in Fig. \ref{gam13even1}. As can be seen in this figure, it involves the bare $\sigma$ vertex as well as the PA $\Gamma^{(2,2)}$ and $\Gamma^{(3,3)}$ 1PI vertices. As it stands, though, we would have to solve the independent equation for $\Gamma^{(3,3)}$ in order to make progress (an analysis of which can be found in Appendix \ref{gam33}, where it is shown that its equation requires numerical methods to be solved). Moreover, this expression is not well suited for the analysis of universal properties, because it is expressed in terms of the bare vertex and not in terms of the full $\Gamma^{(1,3)}$ vertex. Fortunately, the fact that we only deal with PA vertices allows us to find an easier relationship for $\Gamma^{(1,3)}$, using the already known property which allows to find closed forms for PA vertices. Notice that $\Gamma^{(2,2)}$ and $\Gamma^{(3,3)}$ have always two possible closed decompositions, being vertex of the 
form $\Gamma^{(n,n)}$ (see discussion in Section \ref{pa}). Specifically, we can rewrite the equation for $\Gamma^{(1,3)}$ in the form shown diagrammatically in Fig. \ref{gam13even}, which can be written
\begin{multline}
 \Gamma^{(1,3)}(p,\tilde p_1,\tilde p_2,\tilde p_3)=-6\sigma \\
-2\lambda\int_q G(q) G(\tilde p_1+\tilde p_2 -q) \Gamma^{(1,3)}(p,q,\tilde p_1+\tilde p_2-q ,\tilde p_3)\\
-2\lambda\int_q G(q) G(\tilde p_1+\tilde p_3 -q) \Gamma^{(1,3)}(p,q,\tilde p_1+\tilde p_3-q ,\tilde p_2)\\
-2\lambda\int_q G(q) G(\tilde p_2+\tilde p_3 -q) \Gamma^{(1,3)}(p,q,\tilde p_2+\tilde p_3-q ,\tilde p_1)
\end{multline}
The highly symmetric form of this equation suggests the following ansatz for the functional form of $\Gamma^{(1,3)}$ (which can be easily checked using the equality $p=\tilde p_1+\tilde p_2+\tilde p_3$)
\begin{equation}
 \Gamma^{(1,3)}(p,\tilde p_1,\tilde p_2,\tilde p_3)= -2\sigma(p,\tilde p_1)-2\sigma(p,\tilde p_2)-2\sigma(p,\tilde p_3)
\end{equation}
In terms of $\sigma(p,\tilde p)$ the equation becomes
\begin{multline}
 \sigma(p,\tilde p)=\sigma-2\lambda\int_qG(q) G(p-\tilde p-q)\\
  \times \Big(\sigma(p,q)+\sigma(p,p-\tilde p-q)+\sigma(p,\tilde p) \Big)
\end{multline}
Using the known expression for $l(p)$, Eq. (\ref{lambda(p)}), we obtain
\begin{multline}
 \sigma(p,\tilde p)=\frac \sigma \lambda l(p-\tilde p) -2l(p-\tilde p)\int_q G(q) G(p-\tilde p-q)\\
  \times \Big(\sigma(p,q)+\sigma(p,p-\tilde p-q) \Big)
\end{multline}
For the calculation of $d_c$, it is enough to analyze the $p=0$ behavior (we are interested in the IR fixed point structure of the theory). Defining
\begin{equation}
 \sigma(\tilde p)=\sigma(p=0,-\tilde p)
\end{equation}
(notice that we have chosen a minus sign in the definition), and after a change of variables inside the integral, we get
\begin{equation}\label{eqsigma1}
 \sigma(\tilde p)= \frac \sigma \lambda  l(\tilde p) -4l(\tilde p) \int_q G(q) G(\tilde p - q) \sigma(q)
\end{equation}
From now on we will omit the tilde in $\tilde p$ for notational simplicity. The quantity we are mostly interested in is $d_\sigma$, the scaling dimension of $\sigma$ in the IR limit
\begin{equation}
 \sigma(p)\sim |\mathbf{p}|^{d_\sigma} \qquad \mathrm{for}\quad \nu,|\mathbf{p}|^2 \ll \lambda^{\frac{2}{2-d}}
\end{equation}
Naive power counting yields $d_\sigma=2$, which would imply that the branching perturbation $\sigma$ is relevant for all $d$, but fluctuations of course change this value of $d_\sigma$, and could even make it negative, which would imply the irrelevance of the $\sigma$ perturbation.

In order to solve Eq. (\ref{eqsigma1}) it is useful to define the quantity
\begin{equation}
 \hat \sigma (p)=\frac{\sigma(p)}{l(p)}
\end{equation}
whose behavior in the IR is expected to be of the form $\hat \sigma(p)\sim |\mathbf{p}|^{d-d_\sigma}$ (recall that $l(p)\sim |\mathbf{p}|^{2-d}$ in that regime). The equation for $\hat \sigma$ reads
\begin{equation}\label{sigmahat1}
 \hat \sigma(p)=\frac \sigma \lambda -4\int_q G(q) G(p - q) \hat \sigma(q) l(q)
\end{equation}
Using this exact expression and expanding in $\epsilon=2-d$ we recover the 1-loop result $d_c=4/3$, as well as the 2-loop result $d_c\simeq 1.1$ \cite{cardy96}. These results follow from a perturbative series in $\sigma$ and $\lambda$, and from a simultaneous expansion in $\epsilon=2-d$. The details of these calculations can be found in appendix \ref{2loops}.

In order to get an exact result for $d_\sigma$ it is convenient to get rid of the bare reaction rates, as we are interested in the universal IR scaling behavior. Let us start by doing so in the case of $l(p)$, which will be useful in what follows. The IR limit is taken by making $\lambda \to \infty$ (more precisely, by considering $\nu, |\mathbf{p}|^2 \ll \lambda^{2/(2-d)}$, the typical momentum scale set by the bare annihilation rate). This can be done safely for $d<2$, and is a subtle limit when one studies directly $d=2$ in order to perform the $\epsilon$-expansion, as will be further commented in Appendix \ref{2loops}. By exploiting expression (\ref{lambda(p)}) we obtain the IR behavior
\begin{align}
 \left(l^{IR}(p)\right)^{-1}&=2 \int_q G(q) G(p-q)\notag \\
 &=2\int \frac{d^d q}{(2\pi)^d}\int \frac{d\omega}{2\pi} \frac{1}{q^2+i\omega}\frac{1}{(p-q)^2+i(\nu -\omega)}
\end{align}
and thus
\begin{equation}\label{lambdaIR}
 l^{IR}(p)=\frac{(4\pi)^{d/2}}{2^{1-d/2}\Gamma(1-d/2)}\left( \frac{p^2}{2}+i\nu \right)^{1-d/2}
\end{equation}

Now we can return to $\hat \sigma$. As we are only interested in its scaling behavior, it proves convenient to subtract to (\ref{sigmahat1}) its value at zero $\hat \sigma(p=0)$, which is zero in the IR for $d<2$, given that we expect $d_\sigma<d$. This is seen to be true in the $\epsilon$ expansion around $d=2$, and must be true near the sought-for $d_c$, where $d_\sigma$ should be zero. Our results will later confirm $d_\sigma<d$. We thus have
\begin{equation}
\hat \sigma(p)=-4\int_q \hat \sigma(q) l(q) G(q)\Big( G(p - q) -G(-q)\Big) 
\end{equation}
This is a complicated equation, and to be able to solve it, we must take into account the scaling invariance we expect from its solution. We exploit scale invariance in order to define the scaling function $\tilde \sigma(\tilde \nu)$
\begin{align}\label{scaling}
 \hat \sigma(\mathbf{p},\nu)&=|\mathbf{p}|^{d-d_\sigma}\tilde \sigma (\tilde \nu), & \tilde \nu &=\frac{\nu}{|\mathbf{p}|^2}
\end{align}
Observe that we are performing a perturbation around the PA fixed point, whose anomalous dimensions are zero (that is, $\eta=0$, $z=2$, as already mentioned). Accordingly, the natural scaling variable is $\tilde \nu=\nu/|\mathbf{p}|^2$.

We can now write an equation for $\tilde \sigma(\tilde \nu)$, using the form (\ref{lambdaIR}) for $l(p)$ and choosing as variables $\tilde \omega=\omega/q^2$, $\tilde q=|q|/|p|$ and $u=\cos \widehat{(\mathbf{p}, \mathbf {q})}$
\begin{multline}
 \tilde \sigma(\tilde \nu)=-4\left(\frac{(4\pi)^{d/2}}{2^{1-d/2}\Gamma(1-d/2)}\right)\left(\frac{2\pi^{\frac{d-1}{2}}}{(2\pi)^{d+1}\Gamma\left(\frac{d-1}{2}\right)}\right)\\
\times \int_0^\infty d\tilde q\,\tilde q^{d-d_\sigma+1}\int_{-1}^1 du\,(1-u^2)^{\frac{d-3}{2}}\int_{-\infty}^\infty d\tilde \omega \, \tilde\sigma(\tilde \omega) \left(\frac 1 2 +i\tilde \omega\right)^{1-d/2}\\
\times \frac{1}{1+i\tilde\omega} \left( \frac{1}{1+i\tilde \nu +\tilde q^2(1+i\tilde \omega)-2\tilde q u}-\frac{1}{\tilde q^2(1+i\tilde \omega)} \right)
\end{multline}

\begin{center}
\begin{figure}[ht]
\includegraphics[angle=-90,width=0.45\textwidth]{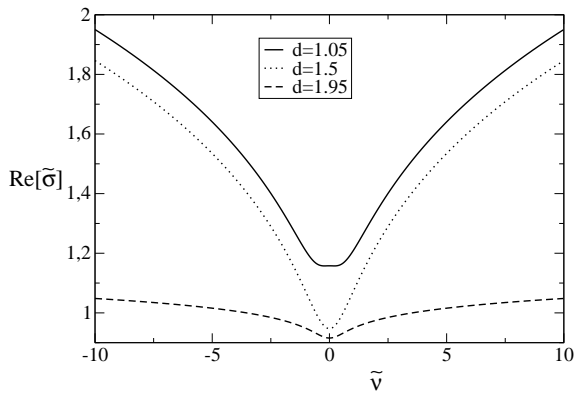}
\caption{$\tilde \nu$-dependence of the real part of the scaling function $\tilde \sigma (\tilde \nu)$ for several values of $d$.}
\label{sigmahat}
\end{figure}
\end{center}
This equation is still too complicated to be solved analytically, and requires a numerical solution. A convenient way to do that is to make an expansion in $u$, which turns out to be rapidly convergent. We then proceed as follows: at each order in the expansion in $u$ we adjust $d_\sigma$ at a given value of $d$, by numerically iterating this equation in order to reach a fixed functional form for $\tilde \sigma (\tilde \nu)$ in a lattice of $N_\nu$ points with a resolution $\delta \nu$. We have checked the convergence in $u$ and in the numerical parameters $\delta \nu$ and $N_\nu$, used for the computation of integrals. This  procedure gives always a converged scaling function $\tilde \sigma (\tilde \nu)$, which confirms a posteriori the scaling form ansatz (\ref{scaling}). In Fig. \ref{sigmahat}, we show the explicit $\tilde \nu$ dependence of the function $\tilde \sigma (\tilde \nu)$ for some values of $d$. As can be seen, it is a non-trivial function of its argument, which may explain the qualitative 
difference between our results and previous approximate results. Observe that LPA and 1-loop analysis are based on a constant coupling $\sigma$ (without dependence on frequency and momentum). As expected, this dependence becomes weaker as $d$ approaches $2$.

\begin{center}
\begin{figure}[ht]
\includegraphics[angle=-90,width=0.45\textwidth]{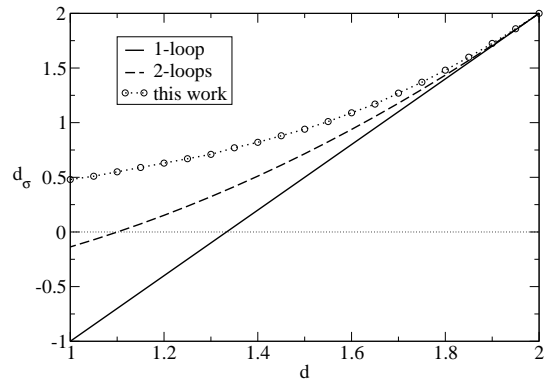}
\caption{Results for $d_\sigma$, showing there is no change in the RG relevance for the branching rate $\sigma$.}
\label{orden6}
\end{figure}
\end{center}
\begin{center}
\begin{figure}[ht]
\includegraphics[angle=0,width=0.45\textwidth]{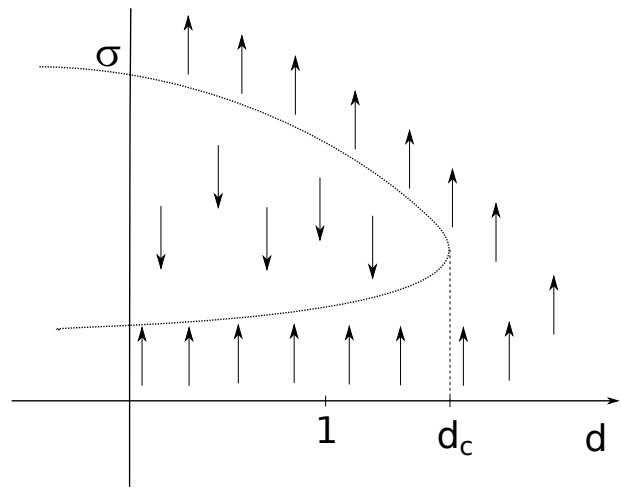}
\caption{Sketch of the relevance of the $\sigma$ perturbation in BARW-PC, compatible with the results in this work and with simulations.}
\label{graph2}
\end{figure}
\end{center}
This procedure allows us to find the value of $d_\sigma$ as a function of $d$, the results of which are plotted, together with previous perturbative results, in Fig. \ref{orden6}. There one can see that even if $d_\sigma$ gets smaller when $d$ decreases, it remains always positive. This is an unexpected result, which deserves a careful discussion. First of all, it is important to observe that this result does not rule out the existence of a new fixed point $F^{PC}$ for small $d$, governing the properties of the PC universality class. A new fixed point can indeed appear, but for a nonzero value of the branching rate $\sigma^*$, as seen for example in the sketched flow shown in Fig. \ref{graph2}. This possible scenario would mean, in particular, that the low branching phase of the model would have a behavior different from PA, governed by a new absorbing fixed point, with as-yet unknown scaling properties. 

A fixed point governing the absorbing phase is needed in order to reproduce the well established power-law like behaviour in the absorbing phase, as seen by Monte-Carlo simulations. Notice that this power-law behaviour would be in this case obtained without any parameter fine-tuning, corresponding to what is usually known as a quasi long range order phase, as observed before for example in equilibrium statistical mechanics in \cite{qlro}. The validity of such an scenario can be studied either by using Monte-Carlo methods or by going to higher orders in the $\sigma$-expansion, or perhaps by means of the NPRG at orders higher than the LPA.

As mentioned before, there exists an exact result in $d=1$ \cite{takayasu92} which seems to indicate that in this universality class $\sigma$ is in fact irrelevant with respect to the PA fixed point in that dimension. We can explain this difference observing that the model used in \cite{takayasu92} is defined with $\lambda=\infty$, and indeed presents no phase transition at all for whatever value of $\sigma$. Now, the IR limit corresponds to $\nu, |\mathbf{p}|^2 \ll \lambda^{2/(2-d)}$, but this does not allow us to take $\sigma=0$ when compared to $\lambda$. Looking at Eq. (\ref{sigmahat1}), $\lambda=\infty$ implies $\hat \sigma\equiv0$, so that the relevant direction corresponding to $\sigma$ is no longer accessible by studying $\sigma$ as a perturbation. This is true for all $d$. Indeed, the results of \cite{tauber05} are also compatible with this scenario: in fact they show an irrelevant $\sigma$ for all $d$ when $\lambda=\infty$. For example, at 
1-loop level (approximation valid close to $d=2$) we have (see appendix \ref{2loops})
\begin{equation}
 \hat \sigma(p) \sim \frac {\sigma} {\lambda^3}  l^2(p)
\end{equation}
so that we see explicitly that $\lambda \to \infty$ yields $\hat \sigma \equiv 0$, and $\hat \sigma$ is no longer associated with the relevant branching direction. This is in evident contradiction with the fact that there is an unstable direction in that dimension. Thus, we think that the exact calculation in \cite{takayasu92} does not apply to BARW-PC, the system in which we are interested.

Also, there exist a result in \cite{vernon} in which branching and annihilating systems of particles performing L\'evy flight dynamics are studied. In it, the authors show that a change in the value $\sigma$ of the L\'{e}vy flight exponent can be made to correspond to a change in the dimension $d$ of the corresponding standard BARW system. This is used to recover $d_c=4/3$ for BARW-PC. The analysis, however, is made by means of a 1-loop perturbative expansion in $\lambda$ and $\sigma$, which explains the coincidence with the results of \cite{cardy96}. A re-analysis of L\'evy flight dynamics can be envisaged within the approach proposed in this work.

In what respects Monte Carlo studies of the low branching regime of this system, they have until now, as far as we know, also been mostly made in the limit $\lambda\to \infty$ \cite{jensen,odorrev} mentioned before. They are compatible with the standard scenario, but within the criticisms previously pointed out.

Let us emphasize that Fig. \ref{graph2} only shows one of the possible scenarios allowing for the compatibility of all what is known about the PC transition. This scenario is not a result of this work, but only what we consider the simplest possibility. Other explanations may well exist , and we do not pretend that Fig. \ref{graph2} is the final word about this issue.

\section{Summary and discussion}

In this work we have applied field theoretical methods to answer some non-trivial questions about a class of reaction-diffusion systems. We have proceeded by exploiting the special case of Pure Annihilation, a system which does not present a phase transition but which nonetheless possesses a non-trivial fixed point in the RG sense. In order to do so, we took advantage of its simple structure, as well as the symmetries and causal properties of the system (which in fact allowed us to go beyond perturbation theory). We have then applied an expansion in the branching rate around Pure Annihilation, giving us access to the small branching regime of BARW, both with and without an additional parity conserving symmetry.

We have chosen to concentrate, as a first order example, in some important properties of these systems, usually very difficult to control but that become possible to solve within the present method. In the case of the system of reactions $2A\to \emptyset$, $A\to2A$, which belongs to the DP universality class, we have given an explicit proof of the existence of a phase transition in all space dimensions, already seen in previous numerical solutions of approximated versions of the NPRG flow equations, and in Monte-Carlo simulations. We have moreover calculated exactly the non-universal threshold value for the annihilation rate in order to find this phase transition in two sample systems. This result is beyond the possibilities of usual perturbation theory.

In BARW-PC, where the parity of the number of particles is conserved, we have concentrated on the value $d_c$ of the upper critical dimension, that was previously believed to be somewhere between $d=1$ and $d=2$. Previous 1-loop and LPA results indicated $d_c \simeq 4/3$. By truncating our equations at one-loop order we were able to recover this approximate result, as well as the two-loop result of \cite{tauber05}. Surprisingly, we have found that the appearance of the PC fixed point associated with $d_c$ must occur at a nonzero value of the branching rate, which would be compatible with a scenario where there exists not one but two new fixed points for $d<d_c$. Further investigation of this issue should be performed, either by a higher order expansion in $\sigma$ or by lattice simulations,  or by the use of the NPRG method at orders higher than the LPA. Work in some of these directions is already underway.

Let us emphasize that the $\sigma$-expansion here introduced represents an expansion around a non-trivial (non-Gaussian) model, which in particular implies, as explained in the text, that the first order results obtained in this work represent the first term in a convergent series. 

The $\sigma$-expansion has allowed us to obtain results not accessible with the usual perturbative expansion, while still being (for the most part) analytical. This kind of ideas could in principle be generalized to other field theoretical systems, and future applications can be thought within the study of out of equilibrium systems. In what concerns BARW, a second order expansion in $\sigma$ would in principle allow for the approximate calculation of critical exponents. Extensions to other out of equilibrium systems can also be envisaged, for example in the study of PCPD, or of the Cole-Hopf version of the KPZ equation \cite{wiese98}, which has a structure very reminiscent of Pure Annihilation.

\acknowledgements

This paper is dedicated to the memory of Mario Wschebor. We want to thank L. Canet for the numerical data of reference \cite{canet2} and M. Wschebor for pointing out a rigorous argument for the large $d$ behaviour of $I(d)$. We also thank M. Tissier, I. Dornic, A. Ran\c{c}on and H. Chat\'e for useful discussions, and Bertrand Delamotte for carefully reading and commenting an earlier version of the manuscript. We acknowledge partial support from the PEDECIBA program and ANII (Grant FCE-2009-2694). F.B. ackowledges support from the CNRS.

\appendix

\section{Study of the $\Gamma^{(3,3)}$ vertex in PA}
\label{gam33}

In this appendix we will study the $\Gamma^{(3,3)}$ vertex in PA with the methods intruduced in Section \ref{pa}. Unfortunately, as we will see, this will not be enough to find a complete analytical solution for the vertex. Such a solution would need a numerical implementation, beyond the scope of the present work.
\begin{center}
\begin{figure}[ht]
\includegraphics[width=0.45\textwidth]{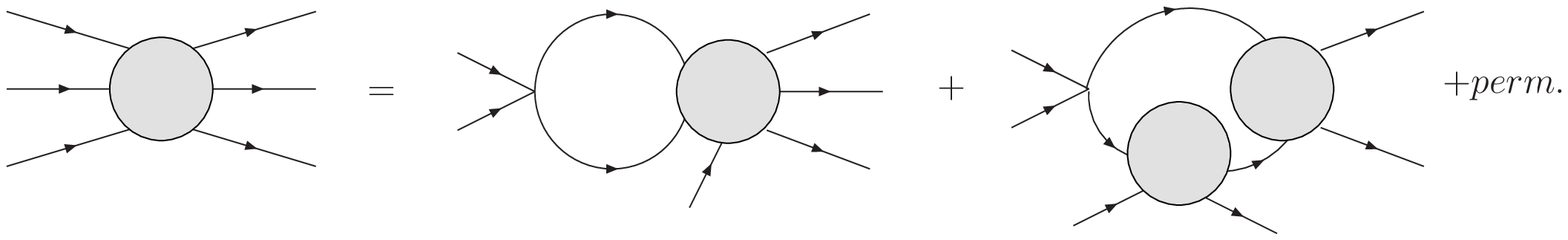}
\caption{Closed equation for $\Gamma^{(3,3)}$ in PA.}
\label{gam33fig}
\end{figure}
\end{center}

As explained in the main text, every diagram for $\Gamma^{(3,3)}$ is of the form shown in Fig. \ref{gamnm}. If we analyze the possible 1PI contributions we end up with the two diagrams shown in Fig. \ref{gam33fig}. Using the known form for $l(p)$ and the corresponding symmetry factors, the equation corresponding to this diagram reads
\begin{multline}
 \Gamma^{(3,3)}(p_1,p_2,p_3,\tilde p_1,\tilde p_2,\tilde p_3)= \\64\lambda \bigg(\int_q G(q)G(p_1+p_2-q) \Big[ l(\tilde p_2+\tilde p_3)G(p_3+q-\tilde p_1) l(q+p_3)\\
+l(\tilde p_1+\tilde p_3)G(p_3+q-\tilde p_2) l(q+p_3) +l(\tilde p_1+\tilde p_2)G(p_3+q-\tilde p_3) l(q+p_3) \Big]\\
+\int_q G(q)G(p_1+p_3-q)l(p_2+q)\Big[G(p_2+q-\tilde p_1)l(\tilde p_2+\tilde p_3) \\
+ G(p_2+q-\tilde p_2)l(\tilde p_1+\tilde p_3) +G(p_2+q-\tilde p_3)l(\tilde p_1+\tilde p_2)\Big]\\
+\int_q G(q)G(p_2+p_3-q)l(p_1+q)\Big[G(p_1+q-\tilde p_1)l(\tilde p_2+\tilde p_3) \\
+ G(p_1+q-\tilde p_2)l(\tilde p_1+\tilde p_3) +G(p_1+q-\tilde p_3)l(\tilde p_1+\tilde p_2)\Big]\bigg)\\
- 2 \lambda \int_q G(q)\bigg( G(p_1+p_2-q)\Gamma^{(3,3)}(q,p_1+p_2-q,p_3,\tilde p_1,\tilde p_2,\tilde p_3)\\ 
+G(p_1+p_3-q)\Gamma^{(3,3)}(q,p_1+p_3-q,p_2,\tilde p_1,\tilde p_2,\tilde p_3)\\
+G(p_2+p_3-q)\Gamma^{(3,3)}(q,p_2+p_3-q,p_1,\tilde p_1,\tilde p_2,\tilde p_3)\bigg)
\end{multline}
This equation shows a very symmetric structure, which suggests the ansatz (easily proven to be correct using momentum conservation)
\begin{multline}
 \Gamma^{(3,3)}(p_1,p_2,p_3,\tilde p_1,\tilde p_2,\tilde p_3)\\=f(p_1+p_2,p_3,\tilde p_1+\tilde p_2,\tilde p_3)+\,\textrm{permutations}
\end{multline}
where in fact the last dependence (in $\tilde p_3$ in the equation) is redundant due to momentum conservation. The corresponding equation for $f$ is
\begin{multline}
 f(p_a,p_b,\tilde p_a,\tilde p_b)=\\64\lambda\,l(\tilde p_a)\int_q G(q) G(p_a-q)G(p_b+q-\tilde p_b)l(p_b+q)\\
-2\lambda \int_q G(q)G(p_a-q)\Big[f(p_a,p_b,\tilde p_a,\tilde p_b)+f(q+p_b,p_a-q,\tilde p_a,\tilde p_b)\\
+f(p_a+p_b-q,q,\tilde p_a,\tilde p_b))\Big]
\end{multline}
and, using the explicit form for $l(p)$ this can be rewritten as
\begin{multline}
 f(p_a,p_b,\tilde p_a,\tilde p_b)=\\ 64l(p_a)l(\tilde p_a)\int_q G(q-p_b) G(q-\tilde p_b)G(p_a+p_b+q)l(q)\\
-4l(p_a) \int_q G(q)G(p_a-q)f(p_a+p_b-q,q,\tilde p_a,\tilde p_b)
\end{multline}
where a change of variables has also been performed inside the integrals.

In order to get rid of the incoming and outgoing dressed $l(p)$ vertices we can define
\begin{equation}
 \chi(p_a,p_b,\tilde p_a,\tilde p_b)=\frac{f(p_a,p_b,\tilde p_a,\tilde p_b)}{l(p_a)\tilde l(p_a)}
\end{equation}
whose equation reads
\begin{multline}\label{eqchi}
 \chi(p_a,p_b,\tilde p_a,\tilde p_b)=64\int_q G(q-p_b) G(q-\tilde p_b)G(p_a+p_b+q)l(q)\\
-4 \int_q G(q)G(p_a-q)l(p_a+p_b-q)\chi(p_a+p_b-q,q,\tilde p_a,\tilde p_b)
\end{multline}

This expression can be further simplified, but in the long run a numerical study of it is unavoidable in order to solve it. Notice also that this expression is not explicitly symmetric with respect to the change $p\to\tilde p$. It is easy though to find such a symmetric expression, by combining this expression with the one obtained by using the equation stemming from the diagrammatic ansatz with the $\lambda$ bare vertex in the outgoing legs of the diagram. This symmetric equation does not turn out to be simpler to solve than (\ref{eqchi}).

\section{Non Perturbative Renormalization Group study of BARW}
\label{nprg}

The Non-Perturbative Renormalization Group \cite{berges02,Ellwanger93,Tetradis94,Morris94b,Morris94c,Wetterich:1992yh} is a general framework for the study of strongly correlated systems, mainly for field theoretical problems. It is basically the modern version of Wilson's initial ideas on the RG. One of the main interests for using the the NPRG is that it enables to devise new approximation schemes, useful to tackle problems where more classical tools are of little or no use. In particular, the NPRG stands out as a natural ground for the study of strongly coupled  systems, and the last years have seen an increasing number of applications of NPRG methods to diverse kinds of problems in many areas of (in general strongly correlated) physics \cite{berges02,Delamotte03,Delamotte:2004zg,BMWpre}, including out of equilibrium systems \cite{Delamotte:ito,BMWkpz,BMWkpz2,canet2,canet3,kopietz}. In this regard, the NPRG is particularly well suited 
for the study of critical scale-free regimes, where renormalization group methods are known to give a simple description of the critical behaviour, and where strong correlations are generally present.
Detailed introductions to the NPRG can be found in the references \cite{berges02,Delamotte:2007pf}, and also in particular when applied to out of equilibrium problems in \cite{Delamotte:ito,canet06,Delamotte:2004zg}.

The NPRG formalism relies on the construction of a sequence of scale-dependent effective models for the model under study, interpolating between the short scale physics at the assumed microscopic scale $\Lambda$ and the full long distance physics when the running scale goes to zero, by averaging over fluctuations at each value of this sliding momentum scale $k$. Instead of working at the level of an effective Hamiltonian, as in the original Wilsonian approach, the NPRG is developed in terms of the flow of effective average actions $\Gamma_k$, which are scale dependent modifications of the 1PI generating functional $\Gamma$, in such a way that $\Gamma_k$ only takes into account fluctuations with characteristic momenta $|q| \gtrsim k$. At the scale $k=\Lambda$, no fluctuation is taken into account and $\Gamma_{\Lambda}$ coincides with the microscopic action $S$, whereas full system information is recovered in the limit $\Gamma_{k\to0} = \Gamma$. The procedure for constructing the effective average action $\
Gamma_k$ consists in the addition of a scale dependent mass-like term which partly freezes out the slow modes. This is achieved by adding a regulator term to the original microscopic action of the form \cite{berges02}
\begin{equation}\label{regterm}
\Delta  S_k[\phi,\bar \phi] =\int_x\big(\phi(x),\bar \phi(x)\big)\cdot \hat R_k(\mathbf{\nabla}^2,\partial_t)\cdot \big(\phi(x),\bar \phi(x)\big).
\end{equation}
In the expression (\ref{regterm}), $\hat R_k$ is the (model dependent) regulator matrix, with matrix elements $\sim R_k$, a cutoff function which  behaves as $R_k(\mathbf q^2,\omega)\sim k^2$ (in Fourier space) for small momenta $|\mathbf q| \lesssim k$, in order to generate a decoupling of the ``slow'' modes, and $R_k$ vanishing for large momenta $|\mathbf q| \gtrsim k$, so that the rapid modes remain almost unaltered.
The scale-dependent generating functionals ${\cal Z}_k[J,\bar J]= \int {\cal D}\phi {\cal D} \bar \phi\, \exp(-  S- \Delta S_k +  \int_x J\,\phi +  \int_x \bar J\,\bar \phi)$ are then used to obtain the effective average action $\Gamma_k$, through the Legendre transform of $\log {\cal Z}_k[J,\bar J]$.

At the core of the method there is an exact functional differential equation, the Wetterich equation \cite{Wetterich:1992yh}, which governs the RG flow of $\Gamma_k$ as a function of the scale $k$. This equation can be cast in a way that makes explicit its one-loop, 1PI structure:
\begin{equation}\label{wetterich}
\partial_k \Gamma_k = \frac 1 2 \tilde\partial_k\mathrm{Tr} \int \frac{d^d\, \mathbf{q}}{(2\pi)^d} \frac{d\omega}{2\pi} 
\log\Big(\hat \Gamma_k^{(2)}+\hat R_k \Big)
\end{equation}
where $\tilde\partial_k$ is the derivative with respect to $k$ applied only to the explicit dependence on $k$ of the regulator profile $R_k$, and $\hat \Gamma_k^{(2)}$ is the matrix of second derivatives of $\Gamma_k$ with respect to the fields. One observes that the right hand side of Eq. (\ref{wetterich}) is similar to the 1-loop expression for the $\Gamma$ functional, but with vertices and propagators extracted from $\Gamma_k+\Delta S_k$ instead of $S$. We make extensive use of this 1-loop structure in the following.

Equation (\ref{wetterich}) cannot in general be solved exactly, and one usually has to perform some approximation in order to extract physical information from it. The most broadly used method is the so-called Derivative Expansion \cite{Tetradis94,Morris94c}, which imposes a particular ansatz for the functional form of $\Gamma_k$ in the small momentum regime (in fact, for critical systems it can be argued that it is valid only in the limit of momenta $|\mathbf q| \lesssim k$ \cite{BMW,BMWlong}). The DE, as well as other possible approximations used in the literature \cite{berges02,BMW,BMWlong}, does not rely on any small coupling parameter, so that the approach remains non-perturbative in essence. It is of course also possible to perform usual perturbative calculations within the NPRG formalism.

Here we assume the existence of a suitable regulator function, and in particular one that does not break the symmetries of the field theory. There is still a great degree of arbitraireness in the selection of such a regulator.

As stated in the main text, causal properties, and in particular the so-called It\^o prescription \cite{vankampen,gardiner} are much less trivial to study in the NPRG formalism when compared with the usual perturbative approach \cite{Delamotte:ito}. One can show that the earliest time in the time dependence of a vertex must correspond to a $\phi$ field, and conversely the last time must correspond to a $\bar \phi$ field. This implies that the 2-point function $\Gamma^{(1,1)}_k$ remains causal for all $k$. The proof of this causal property will be shown in appendix \ref{causal}.

In the particular case of PA, it will be shown in Appendixes \ref{SD} and \ref{causal} that this property is generalized to 3 first-in-time $\phi$ legs and 3 last-in-time $\bar \phi$ legs for any $\Gamma_k^{(n,n)}$.  This implies in particular that the 4-point function $\Gamma^{(2,2)}_k$ and the 6-point function $\Gamma^{(3,3)}_k$ are completely causal, in the sense that all its incoming legs must correspond to times smaller than those associated with any of its outgoing legs. As we shall see below, this property greatly simplifies the study of vertices with a low number of legs.

\section{Non-perturbative proof of the form of the closed equation for $\Gamma^{(n,m)}$ in PA}
\label{SD}

In this Appendix we will use the NPRG as discussed in Appendix \ref{nprg} in order to show the general diagrammatic property used in the main text (see Fig. \ref{gamnm}), which relates $\Gamma^{(n,m)}$ with $\lambda$ and vertices of at most $(n,m)$ legs. This relation has been presented in the main text at all orders in a perturbative expansion in $\lambda$, but here we sketch a non-perturbative proof for it. 

First, observe that we can write the property we want to show in terms of a specific form for $\Gamma_k$, the average effective action for the theory at any scale $k$
\begin{equation}
 \Gamma_k=  S^{PA}+\int_x \phi^2(x)\int_{x',x''}G_k(x'-x)G_k(x''-x)\tilde \Gamma_k[x',x'',\phi,\bar \phi]
\end{equation}
with
\begin{align}
 \frac{\delta \tilde \Gamma_k[x',x'',\phi,\bar \phi]}{\delta \phi(x)} &= 0 & \mathrm{if}\, t<\mathrm{max}(t',t'')\notag \\
\frac{\delta \tilde \Gamma_k[x',x'',\phi,\bar \phi]}{\delta \bar \phi(x)} &= 0 & \mathrm{if}\, t<\mathrm{max}(t',t'')\label{ansatz}
\end{align}
indeed, this form ensures that any $\Gamma_k^{(n,m)}$ vertex will consit in a series of bare terms given by the action $S^{PA}$, together with renormalized terms which always begin (temporally speaking) by a bare $(2,2)$ vertex.

The property we want to show is obviously true at the bare level, with $\tilde \Gamma_\Lambda= 0$. Now we proceed in an iterative way, assuming that the property we want to prove is true at a RG scale $k_0$, and checking that it continues to be valid for a scale $k_0-\delta k$. 

By hypothesis then, we take that at $k=k_0$ all vertices $\Gamma^{(n,m)}_{k_0}$ can be decomposed as a diagram containing a $\lambda_4$ bare vertex (with two simultaneous incoming $\phi$ legs) and a $\tilde \Gamma^{(n,m)}_{k_0}$, this last function being constrained by the condition that the full diagram must be 1PI. As explained before, the NPRG equations for any vertex $\Gamma_k^{(n,m)}$ can be represented diagramatically by 1-loop diagrams, where vertices and propagators are read from $\Gamma_k+\Delta S_k$. To those diagrams one must apply the operator $\tilde \partial_k$ in order to
obtain $\partial_k \Gamma_k^{(n,m)}$. Each one of these terms will consist in a number of $\Gamma_k^{(l,s)}$ vertices joined together by an internal loop of propagators. We can distinguish between internal lines, pertaining to the internal loop, and external lines. 

Consider then a typical diagram contributing to $\partial_k\Gamma_k^{(n,m)}(t_1,\dots,t_n,\tilde t_1,\dots,\tilde t_m)$, where we emphasize the time dependence of the vertices. We now define as $t_0$ the smallest time for any incoming external leg, $t_0 \leq t_k$ $\forall k$. Its corresponding leg is attached to one of the vertices in the loop, and by hypothesis at $k=k_0$, the two smallest times in this vertex must correspond to an incoming bare $\lambda_4$ vertex. We have then two possibilities to consider:

1) Both $t_0$ legs are external legs. This means that $\partial_s \Gamma_k^{(n,m)}$ can also be decomposed in the form (\ref{ansatz}), which implies the desired property for $k=k_0 - \delta k$

2) Only one $t_0$ leg is an external legs, and the other one is an internal incoming leg. Thus, given the non-renormalization of the propagators, there is another vertex with an outgoing leg at a time previous to $t_0$. But at $k=k_0$ this vertex should have at least two incoming legs with a corresponding time previous to $t_0$. At least one of these incoming legs must be external, contradicting our assumption.

The desired property is then preserved all along the NPRG flow, showing that it is a fully nonperturbative result as stated in the main text.

\section{Causal properties in BARW and PA within the NPRG}
\label{causal}

In this appendix we will describe the non-perturbative causal properties in BARW and PA. Altough these properties have a simple expression in terms of a perturbative expansion \cite{cardy96}, they are somewhat difficult to prove in the context of the NPRG \cite{Delamotte:ito}, which is what we choose to do here, in order to have non perturbative versions of the results. 

The method used here is completely analogous to that used in appendix \ref{SD} in order to prove the diagrammatic property used in the main text. In fact, this property can be seen as a causal property of PA. Here we show that it can be slightly generalized. But before that let us start with a simple causal property for generic BARW.
\begin{figure}[ht]
\includegraphics[width=0.3\textwidth]{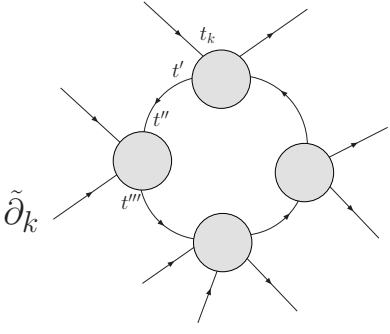}
\caption{An example of a diagram contributing to $\partial_k \Gamma_k^{(n,m)}$ in the NPRG for a generic theory, see text.}
\label{Fig1}
\end{figure}

Let's concentrate on theories where at bare level, all vertices verify that in the time domain
\begin{equation}
\Gamma_\Lambda^{(n,m)}(t_1,\dots,t_n,\tilde t_1,\dots,\tilde t_m)=0
\end{equation}
if one of the times in the set $\{t_1,\dots,t_n\}$ is strictly larger than all times in the set $\{\tilde t_1,\dots,\tilde t_m\}$. For the $2$ point functions this is equivalent to imposing the It\^o prescription \cite{vankampen,Delamotte:ito} by requiring that the propagator in time is strictly causal ($G(t,t'=t)=0$). For all other correlation functions this can be seen as a consequence of the locality of the bare theory. These properties apply for all BARW, as can be checked by looking at Eq. (\ref{actiongeneral}). We are going to prove now that if these properties take place at bare level they remain true all along the flow. This implies, in particular, that all $\Gamma^{(0,n)}_k$ are zero, due to the non-existence of \emph{any} incoming time in that case.

As explained before, the NPRG equations for any vertex $\Gamma_k^{(n,m)}$ can be represented diagramatically by 1-loop diagrams, where vertices and propagators are read from $\Gamma_k+\Delta S_k$. To those diagrams one must apply the operator $\tilde \partial_k$ in order to
obtain $\partial_k \Gamma_k^{(n,m)}$. We represent terms of these equations before the application of the $\tilde \partial_k$ operator. Each one of these terms will consist in a number of $\Gamma_k^{(s,t)}$ vertices joined together by an internal loop of propagators.

Assuming that the property we want to prove is true at a RG scale $k_0$, we will check that it continues to be valid for a scale $k_0-\delta k$, which implies the result given that by hypothesis it is valid at the bare level. Consider then a typical diagram contributing to $\partial_k\Gamma_k^{(n,m)}(t_1,\dots,t_n,\tilde t_1,\dots,\tilde t_m)$ as shown in Fig. \ref{Fig1}, where we have chosen to explicit the time dependence in each vertex. Suppose now that $\partial_k\Gamma_k^{(n,m)}(t_1,\dots,t_n,\tilde t_1,\dots,\tilde t_m)$ has a non zero contribution with a given $t_k$ strictly larger than all $\tilde t_{k'}$. Let us consider the largest such time $t_k$. Its corresponding leg is attached to one of the vertices in the loop, and by hypothesis, at $k=k_0$, the largest time in this vertex must correspond to an outgoing leg. However, this leg cannot be an external leg, because we have assumed that $t_k$ is the largest external time.
Accordingly, this time (wich we will call $t'$, see Fig. \ref{Fig1}) must be in an outgoing
internal line. This line arrives to another vertex at a time $t''>t'>t_k$
that consequently is strictly larger than any external time, and that is
associated to an incoming internal line. Again, by hypothesis,
there must be an outgoing leg with a time $t'''\geq t''$ but this leg cannot be
an external leg (because $t''>t_k$). Consequently $t'''$ must correspond to
the other internal line that must consequently be outgoing.
Following the internal (one loop) line and repeating the reasoning for
each vertex, one returns to the initial vertex at a time strictly larger
than the supposedly largest time. This is absurd, proving the desired property.

Now, in the particular case of PA this can be further generalized. In fact, in PA we will have that for any $\Gamma_k^{(n,m)}$ the first \emph{three} times in the vertex must correspond to incoming $\phi$ legs (when possible, that is, when $n>2$). This follows in a similar way: the property is trivial in PA at the bare level, and we will assume by hypothesis that it is true at the scale $k_0$. Again, consider a typical diagram contributing to
$\partial_k\Gamma_k^{(n,m)}(t_1,\dots,t_n,\tilde t_1,\dots,\tilde t_m)$. By the property shown in appendix \ref{SD}, we know that the two smallest external times $t_1$ and $t_2$ correspond to two incoming legs in the same vertex in the 1PI loop. 

Let us first assume that the third time $t_3$ is also attached to this same vertex. By hypothesis, at $k_0$, it must correspond to an incoming leg. If it is an external leg the result follows. If it is an internal loop leg, then it must be an outgoing leg from another vertex in the loop. But again by hypothesis at $k_0$, there must be an external incoming leg in this other vertex with a corresponding time smaller than $t_3$, which is absurd (unless $n<3$, which would correspond to a particular case). 

The other possibility is that the third smallest time $t_3$ is attached to another vertex. If this time corresponds to an outgoing leg, we will have that in order for this vertex not to have an incoming external leg (which would by hypothesis correspond to a time smaller than $t_3$) then this must be a $\Gamma_k^{(2,2)}$ vertex with both its incoming lines being internal loop lines. But then the same argument as before applies, and we can find an smaller time in yet another adjacent vertex. This completes the proof.

\section{NPRG equations for the simplest vertices}
\label{nprgvertex}
\begin{center}
\begin{figure}[ht]
\includegraphics[width=0.3\textwidth]{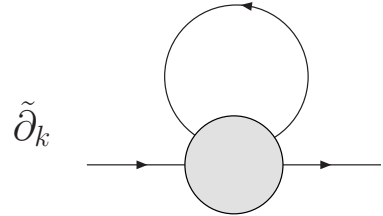}
\caption{Tadpole diagram which would contribute to $\partial_k \Gamma_k^{(1,1)}$ forbidden by It\^o.}
\label{fig3}
\end{figure}
\end{center}
The first consequence to be taken from the causality results in the previous appendix is the non-renormalization of the 2-point function $\Gamma_k^{(1,1)}$. As can be easily seen, the tadpole diagram (that in this case includes $\Gamma_k^{(2,2)}$, see Fig. \ref{fig3}) is zero due to these causal properties: in order to exist, one of its outgoing legs must be associated with a time $\tilde t_0$ smaller than one of the incoming legs, which is impossible in PA. The remaining possible diagram would include a $\Gamma_k^{(1,2)}$ vertex and is accordingly zero. Notice also that the other $2$-point functions $\Gamma_k^{(0,2)}$ and $\Gamma_k^{(2,0)}$ remain equal to zero in the shifted version (Eq. \ref{S_PA}) of PA, in the first case because of property (\ref{propPA}), and in the second case due to causality (see appendix \ref{causal}).

We now explicitly solve the flow equations for the functions $\Gamma_k^{(2,1)}$ and $\Gamma_k^{(2,2)}$. The flow equation for $\Gamma_k^{(2,2)}$ reads
\begin{multline}
 \partial_k \Gamma_k^{(2,2)}(p_1,p_2,p_3,p_4)= \int_{q}\partial_k R_k(q)  G^2_k(q) G_k(p_1+p_2-q)\\
\times \Gamma^{(2,2)}_k(p_1,p_2,-q,-p_1-p_2+q) \Gamma^{(2,2)}_k(q,p_1+p_2-q,-p_3,-p_4)
\end{multline}
which is a closed form equation, given that the other possible term is a tadpole with a $\Gamma^{(3,3)}$ vertex evaluated at an uncausal configuration. This equation is not only closed but, moreover, we can show that it implies that $ \Gamma_k^{(2,2)}(p_1,p_2,p_3,p_4)$ is a function of $p_1+p_2$ only. This is trivially the case in the microscopic initial scale, where $\Gamma^{(2,2)}_{\Lambda}(p_1,p_2,p_3,p_4)= 4\lambda$. Now, if this property is true for all $k\geq k_0$, we can define
\begin{equation}\label{lambda}
l_{k_0}(p_1+p_2)=\frac{1}{4} \Gamma^{(2,2)}_{k_0}(p_1,p_2,p_3,p_4)
\end{equation}
and at this scale the flow equation becomes
\begin{multline}
\partial_k \Gamma_k^{(2,2)}\Big\vert_{k=k_0}(p_1,p_2,p_3,p_4) = \\16 l_{k_0}^2(p_1+p_2) \int_q \partial_k R_k(q)\Big\vert_{k=k_0} G_{k_0}^2(q) G_{k_0}(p_1+p_2-q)
\end{multline}
showing by iteration that $\Gamma^{(2,2)}_k(p_1,p_2,p_3,p_4)$ remains a function of $p_1+p_2$ for any $k$. We then generalize $l_k(p_1+p_2)$ as in (\ref{lambda}) to any scale $k$. This definition allows one to arrive to the simpler flow equation
\begin{align}\label{flow_lambda}
 \partial_k l_k (p) &= 4 l_k^2(p) \int_q \partial_k R_k(q) G_k^2(q) G_k(p-q)
\end{align}
which can be re-written as
\begin{equation}
 \partial_k (l_k^{-1}(p))= 2 \partial_k \int_q G_k(q) G_k(p-q)
\end{equation}
with solution
\begin{equation}
\label{sollambda}
 l_k(p) = \frac{\lambda}{1+2\lambda\int_q \Big( G_k(q)G_k(p-q) - G_\Lambda(q)G_\Lambda(p-q)\Big)}
\end{equation}
Here the dependence on the scale $k$ of the propagator $G_k$ enters only through the regulator function $R_k$. This is completely equivalent to the results in the main text for $k=0$. The general NPRG structure is very reminiscent of the case of the $O(N)$ scalar field theory in the large $N$ limit, where one can also show \cite{D'Attanasio97,BMW} that the NPRG equations can be solved exactly, on similar lines.

As for the three point function, $\Gamma_k^{(2,1)}$ one can start in an analogous way from its flow equation
\begin{multline}
 \partial_k \Gamma_k^{(2,1)}(p_1,p_2,p_3)=\int_q \partial_k R_k(q) G_k^2(q) G_k(p_1+p_2-q) \\
 \times \Gamma^{(2,2)}_k(p_1,p_2,-q,-p_1 - p_2 -q) \Gamma_k^{(2,1)}(q,p_1+p_2-q,-p_1-p_2).
\end{multline}
Given the knowledge of $\Gamma_k^{(2,2)}$ the equation is again closed. Moreover, one can now perform the same type of analysis that has been done for $\Gamma_k^{(2,2)}$ in order to prove that $\Gamma_k^{(2,1)}(p_1,p_2,p_3)$ is a function of $p_1+p_2$ and define
\begin{equation}
 g_k(p_1+p_2)=\frac{1}{2}\Gamma_k^{(2,1)}(p_1,p_2,p_3)
\end{equation}
whose equation reads
\begin{equation}
 \partial_k g_k(p)=4 g_k(p) \lambda_k(p) \int_q \partial_k R_k(q) G_k^2(q) G_k(p-q).
\end{equation}
Then, by using (\ref{flow_lambda}), one deduces that for all $k$
\begin{equation}
 \partial_k \Big(\frac{g_k(p)}{l_k(p)} \Big) = \frac{\partial_k g_k(p)}{l_k(p)}
-\frac{g_k(p) \partial_k \lambda_k (p)}{l_k^2(p)} = 0.
\end{equation}
As $g_\Lambda(p)=2l=2l_\Lambda(p)$, one concludes that
\begin{equation}\label{g=2lambda}
 g_k(p)=2l_k(p)
\end{equation}
with the propagator defined as the regulated bare propagator, which in the case of a continuum theory would read
\begin{equation}
 G_k(q,\omega)=\frac{1}{\mathbf{ q}^2+i\omega+R_k(q)}
\end{equation}

All these results are strictly equivalent to those given in Section \ref{pa} of the main text in the limit $k\to 0$.

Within this formalism we can also recover the results given in Section \ref{odd} of the main text with respect to the threshold of the active-to-absorbing transition, by following similar lines as before.

Notice tough that this NPRG approach is not sufficient to construct a $\sigma$-expansion with closed equations. Indeed, to be able to do that we should have closed expressions for any vertex in PA, in order to re-write vertices in BARW-DP as an expansion in terms of PA vertices. However, the causal properties which we used to ignore the tadpole diagrams in the flow equations for $\Gamma_k^{(1,1)}$, $\Gamma_k^{(2,1)}$ and $\Gamma_k^{(2,2)}$ are not general enough to allow us to close the flow equations for vertices with a higher number of legs. This is problematic even at first order in $\sigma$ in the BARW-PC case, so that we are not able to recover our results for $d_\sigma$ within the pure NPRG formalism.

\section{Scaling dimension $d_\sigma$ at 1- and 2-loop order}
\label{2loops}

In this appendix we show how to recover the perturbative results of \cite{cardy96} for $d_\sigma$ in BARW-PC within our formalism. Let us begin by recalling from the main text the equation for $\hat \sigma$, which reads
\begin{equation}
 \hat \sigma(p)=\frac \sigma \lambda -4\int_q G(q) G(p - q) \hat \sigma(q) l(q)
\end{equation}

Using this expression we can recover the 1-loop and 2-loop results for $d_\sigma$ (and hence $d_c$). In order to do so it is convenient to get rid of the bare level dependence on $\lambda$ and $\sigma$ by writting some sort of RG flow equation. The easiest way to do this is by performing a logarithmic derivative w.r.t. $\nu$, which yields
\begin{align}\label{flowsigma}
 \nu \partial_\nu \hat \sigma(p)=&-4\nu\partial_\nu\int_q \hat \sigma(q) l(q) G(p-q)G(q)\notag \\
=&4 i \nu \int_q \hat \sigma(q) l(q) G^2(p-q)G(q)
\end{align}
This equation can be compared with the corresponding RG equation for $l(p)$, obtained by differentiating Eq. (\ref{lambda(p)})
\begin{align}\label{flowlambda}
 \nu \partial_\nu l(p)=-2l^2(p) \nu\partial_\nu \int_q G(p-q)G(q)
\end{align}

At 1-loop order the $q$-dependence in $\hat \sigma$ should be weak. In the IR one expects this dependence to be dominated by the external momentum $p$, given that the momentum integral is regular for the values of $d$ we are interested in
\begin{equation}
 \nu \partial_\nu \hat \sigma(p)\simeq-4\hat \sigma(p) l(p)\nu\partial_\nu \int_q  G(p-q)G(q)
\end{equation}
which, together with (\ref{flowlambda}) yields
\begin{equation}
 \nu\partial_\nu\left(\frac{\hat \sigma(p)}{l^2(p)}\right)=0
\end{equation}
Accordingly, and given that
\begin{equation}
 \frac{\hat \sigma (p)}{l^2(p)}\xrightarrow{\nu\to\infty} \frac{\sigma}{\lambda^3}
\end{equation}
we have the result
\begin{equation}
 \hat \sigma(p) \sim \frac {\sigma} {\lambda^3}  l^2(p)
\end{equation}
This behaves, when $\nu, |\mathbf{p}|^2\ll \lambda^{2/(2-d)}$, as
\begin{equation}
 \hat \sigma(p) \sim |\mathbf{p}|^{2(2-d)}
\end{equation}
Given the definition of $d_\sigma$ (see Section \ref{even}) we find
\begin{equation}
 d_\sigma=3d-4=2-3\epsilon
\end{equation}
With this expression we find that $d_\sigma$ changes sign at $d_c=4/3$ as expected.

Going now to 2-loop order it is convenient to use the logarithmic derivative of $\hat \sigma$
\begin{equation}
\nu \partial_\nu \log \hat \sigma(p)= -4i\int_q \frac{l(q) \hat \sigma(q)}{\hat \sigma(p)} G^2(p-q)G(q)
\end{equation}
We now introduce in the r.h.s. of this equation the 1-loop result $\hat \sigma(p)\sim l^2(p)$, to obtain
\begin{equation}
\nu \partial_\nu \log \hat \sigma(p)= -4i\int_q \frac{l^3(q)}{l^2(p)} G^2(p-q)G(q)
\end{equation}

We also need the scaling form for $l(p)$, as given by (\ref{lambdaIR}) in the main text. At this point it is enough in order to obtain $d_\sigma$ to restrict to $\mathbf{p}=0$ (given that $\hat \sigma (\mathbf{p},\nu)\sim \nu^{(d-d_\sigma)/2}$)
\begin{multline}
 d-d_\sigma =-8i\nu \frac{(4\pi)^{d/2}2^{-\epsilon/2}}{\Gamma\left(\frac{\epsilon}{2}\right)} \\ \times\int_q \frac{\left(\frac{\mathbf{ q}^2}{2} +i\omega \right)^{3\epsilon/2}}{\left(i\nu \right)^\epsilon}G(\mathbf{q},\omega)G^2(-\mathbf{ q},\nu-\omega)
\end{multline}
which can be evaluated to yield
\begin{equation}
 d_\sigma=2-3\epsilon+3\log\left(\frac 4 3 \right)\epsilon^2
\end{equation}
the known 2-loops result \cite{cardy96}, which corresponds to $d_c= \simeq 1.1$.



\begin{thebibliography}{10}
\bibitem{note1}
In order to prove this property in a mathematically rigorous way, one needs
to prove that if $z_n=1/n \sum_n cos q_n$,then 
$\int dq_1 \dots dq_n (1/(1-z_n))^\alpha\, (*)$ 
is bounded by a constant independent
of $n$, with $\alpha>1$. This would show that the sequence $1/(1-z_n)$ is
uniformly integrable and that $\int dq_1 \dots dq_n 1/(1-z_n) \to 1$.
In order to find a bound for $(*)$ one can use that
$\int dq_1 \dots dq_n (1/(1-z_n))^\alpha=\int_0^\infty dx \mathrm{Prob}(
(1/(1-z_n))^\alpha>x) \leq x_0 + \int_{x_0}^\infty dx \mathrm{Prob}(
z_n>1-x^{-1/\alpha})\, (**)$
where the probability is taken in the sense of considering the $q_n$ as random
variables uniformly distributed in the interval $[-\pi,\pi]$.
In order to give a bound of the second integral in $(**)$, one can use 
standard techniques of large deviations bounds in order to prove that
$\mathrm{Prob}(z_n>1-x^{-1/\alpha})< 1/x^{nC/\alpha}$ with
a constant $C$ independent of $n$. This gives a bounded integral for
$n>\alpha/C$, proving the desired result.
\end{thebibliography}
\end{document}